\documentclass[runningheads,natbib]{svjour3}
\usepackage{graphicx}
\usepackage{lscape}
\newcommand{\solphys}{Solar Phys }
\newcommand{\apj}{ApJ }
\newcommand{\apjl}{ApJL }

\newcommand{\aap}{A\&A }
\newcommand{\apjs}{ApJS }

\newcommand{\araa}{ARAA }
\newcommand{\jgr}{JGR }
\newcommand{\grl}{GRL }

\newcommand{\aapr}{A\&A Revs }
\newcommand{\pasj}{PASJ }

\newcommand{\planss}{Planetary \& Space Science}
\newcommand{\nat}{Nature }
\newcommand{\lssim} {\, \lower3pt\hbox{$\sim$}\llap{\raise2pt\hbox{$<$}}\,}
\newcommand{\gtsim} {\, \lower3pt\hbox{$\sim$}\llap{\raise2pt\hbox{$>$}}\,}
\newcommand{\ssr}{Space Science Revs}

\journalname{Space Science Reviews}

\begin{document}

\title{Structures in the outer solar atmosphere}



\author{ L. Fletcher \and P. J. Cargill \and S. K. Antiochos \and B. V. Gudiksen
}

\authorrunning{L. Fletcher et al.} 

\institute{L. Fletcher \at
              SUPA School of Physics and Astronomy,\
              University of Glasgow,\
              Glasgow,\
              G12 8QQ, \ UK 
              Tel.: +44 141 330 5311\\
              \email{lyndsay.fletcher@glasgow.ac.uk}       
           \and
           P. J. Cargill \at
             Space and Atmospheric Physics,\
             The Blackett Laboratory,\
             Imperial Colege,\
             London,\
             SW7 2BZ, \ UK  \ and\
             School of Mathematics and Staistics,\
             University of St Andrews,\
             St Andrews,\ Fife,\ KY16 9SS,\ UK \\      
            \and
            S. K. Antiochos \at
            NASA Goddard Space Flight Center,\
             Greenbelt,\ MD,\ 20771,\ USA  \\
             \and
             B. V. Gudiksen \at
             Institute of Theoretical Astrophysics, \
             University of Oslo, \ P.O. Box 1029 \ Blindern,\ NO-0315\ Oslo,\ Norway\
             and \ 
             Center of Mathematics for Applications, \
             University of Oslo, \ P.O. Box 1053 \ Blindern, \ NO-0316 \ Oslo, \ Norway    
}

\maketitle

\begin{abstract}
The structure and dynamics of the outer solar atmosphere are reviewed with emphasis on the role played by the magnetic field. Contemporary observations that focus on high resolution imaging over a range of temperatures, as well as UV, EUV and hard X-ray spectroscopy, demonstrate the presence of a vast range of temporal and spatial scales, mass motions, and particle energies present. By focussing on recent developments in the chromosphere, corona and solar wind, it is shown that small scale processes, in particular magnetic reconnection, play a central role in determining the large-scale structure and properties of all regions. This coupling of scales is central to understanding the atmosphere, yet poses formidable challenges for theoretical models.
\end{abstract}


\section{Introduction}

The structure and dynamics of the outer solar atmosphere, defined as extending from the base of the chromosphere into interplanetary space, is a topic of central importance for understanding solar activity. The fundamental questions have been well known for decades: why do the chromosphere and corona have temperatures considerably in excess of the solar photosphere and why do dynamic phenomena such as flares and coronal mass ejections (CMEs) occur? Only the solar magnetic field can provide the required energy to account for these phenomena. 

\begin{figure}
\centering
\hbox{
\includegraphics[height=6cm]{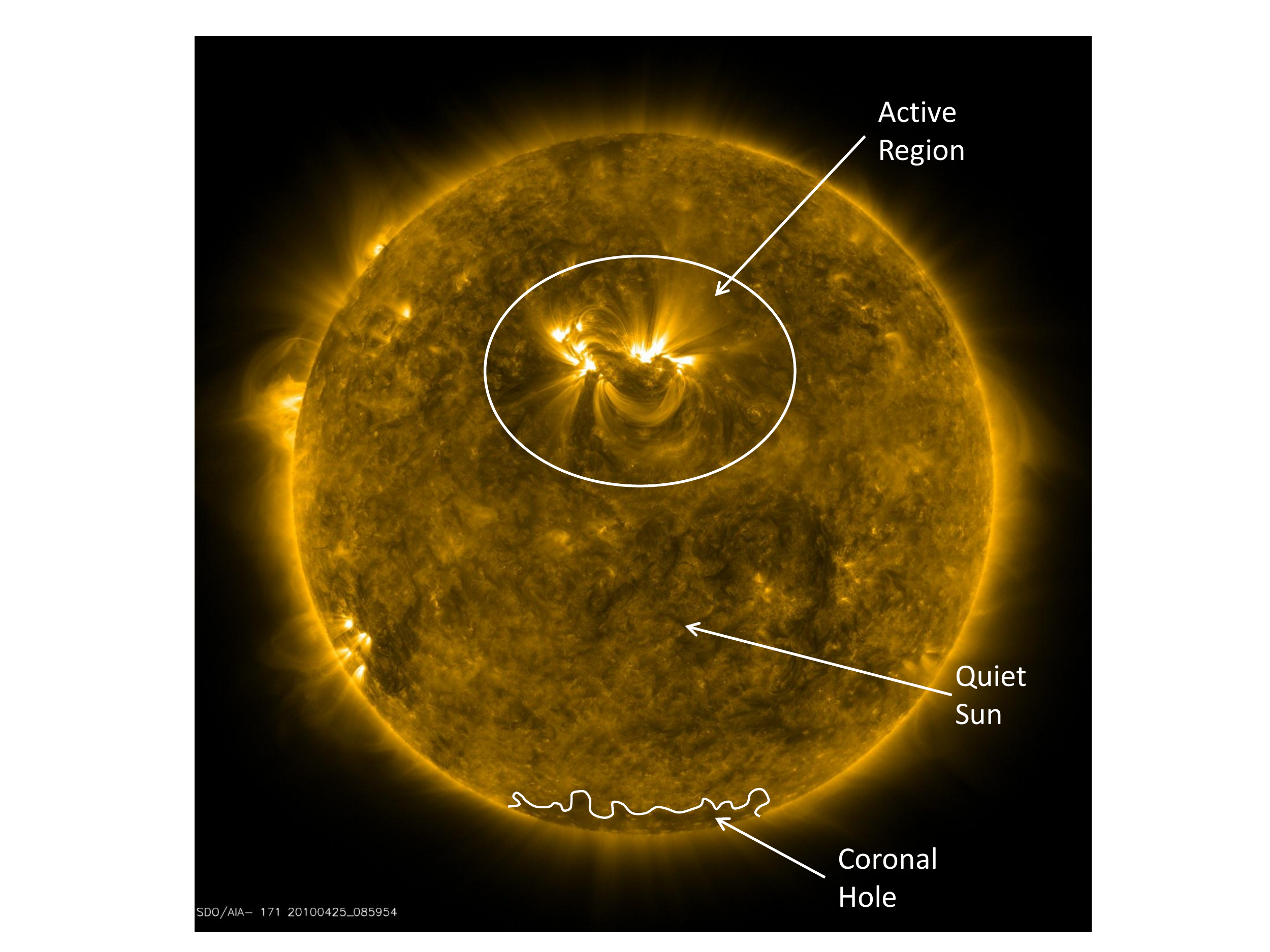}
}
\caption{An image from SDO (adapted from \cite{2011Natur.475..463C}), showing Fe IX emission at around 1 MK. The main features are the dark regions (magnetically open coronal holes: Section 4), bright magnetically closed active regions associated with sunspot groups, brightenings associated with small bipolar field regions (X-ray bright points) and ``the rest": the quiet Sun.}\label{fig:sdo_overview}
\end{figure}

Since the results from the Skylab observatory became available four decades ago, it has been recognised that, as a consequence of the magnetic field, the chromosphere / corona system is highly structured. The chromosphere is a complicated region that extends above the photosphere to a temperature of around 30,000 K and in which the density falls to of order $10^{11} - 10^{12} ~ {\rm cm}^{-3}$ (Section 2). Above this is the corona, with a temperature $>$ 1 MK, considered to be either magnetically closed with field lines forming confined structures usually referred to as coronal loops (Sections 3 and 4), or magnetically open (the source of high-speed solar wind) with field lines extending into interplanetary space (Section 5). There are a variety of magnetically closed structures, the most conspicuous being bright active regions (ARs), the site of intense EUV and X-ray emission. The rest of the magnetically closed corona is termed the quiet Sun, having a smaller level of emission. Figure~\ref{fig:sdo_overview} shows a typical coronal image with the different regions highlighted.

At the photosphere, the Sun's vector magnetic field can be measured routinely, and comprises many discrete sources ranging from small flux elements ($10^{17}$ Mx with a scale of 100 km) to large sunspots ($10^{22}$ Mx) that form a power law distribution over several decades \citep{2011SoPh..269...13T}. The photospheric field strength is roughly its equipartion value, between one and two kG. The relative importance of the magnetic field as one rises through the outer atmosphere is seen by considering the value of the plasma beta ($\beta = 8 \pi p/B^2$). At the photosphere, $\beta$ is of order unity. Above the photosphere the pressure falls rapidly due to the small pressure scale height and so the photospheric field elements expand to fill all space in the upper chromosphere and corona. [Note that at this time, chromospheric field measurements are not made routinely \citep{2004A&A...414.1109L,2005ApJ...623L..53M,2012SoPh..278..471L}.] A consequence is that $\beta$ decreases throughout the chromosphere and becomes $<< 1$, perhaps 0.1 in quiet regions and 0.01 in regions of very strong field, until well out in the corona ($> 2 R_s$) when the expanding solar wind becomes important. This implies that between the middle chromosphere and outer corona, magnetic forces dominate over pressure ones. Plasma also moves with the magnetic field due to magnetic flux freezing, so that a convenient description of these regions is in terms of magnetic field topology and the motion of magnetic field lines. A further consequence of this magnetic connectivity is that any complexity in the photospheric field due to the hierarchy of structures and associated motions of the solar surface must map in some way into the corona. 

One difficulty is the interconnection of the various regions (for example coronal plasma must originate in the chromosphere), and the small scales (sub-resolution) required for the relevant processes that release magnetic energy. In all regions under discussion, effective Reynolds numbers are large, whether based on viscosity, Coulomb or anomalous resistivity (including Hall effects), or Pedersen conductivity, implying small scales. In particular, the magnetic reconnection process that is widely believed to be an essential mechanism for magnetic energy dissipation, and is the subject of much of this paper, requires scales $\ll 1$ km, yet has consequences on global scales which CAN be observed. This poses major challenges, especially for theoretical modelling. 

In keeping with the cross-disciplinary nature of the workshop, this paper is aimed at the non-specialist. It is not a comprehensive review of the subject, rather it focusses on specific aspects in which considerable progress is being made at this time. The structure is traditional, beginning with the chromosphere, then the closed corona, and dynamic phenomenon therein, and leading on to the solar wind. 

\section{The chromosphere}

In the past the solar chromosphere was seen as a thin layer in the solar atmosphere with relatively little influence, except for the formation of three prominent spectral lines: Ca II H,K and $H_\alpha$, and the last of these gave the chromosphere its name, as it is largely $H_\alpha$ that produces the red ring visible in a full solar eclipse. This view of the chromosphere is that it is a layer of the solar atmosphere, roughly 0.5-1.5 Mm thick, with a temperature between 2500 and 25000 K. It is sandwiched between the photosphere and corona, with another thin layer of only a few hundred kilometers, called the transition region (Section 3), separating chromosphere and the corona.
	
However, increasingly sophisticated ground- and space-based observations have made it clear that the chromosphere is not a flat thin layer of constant depth, but rather a highly time-varying undulating layer of varying thickness. Indeed a precise definition of the chromosphere is challenging. The original definition in terms of $H_\alpha$ is not consistent with other definitions that use, for example, a specific temperature range. Such a temperature definition was often based on static one-dimensional calculations of an atmosphere where the aim was to reproduce the intensity of as many spectral lines as possible. One dimensional planar models of the chromosphere are now widely thought to be inadequate, so a present view is that the chromosphere is a highly dynamic layer where many important transitions in physical quantities and processes happen.

\subsection{Structure}

The structure of the chromosphere is determined by the dynamics of the photosphere (whose scales are also present in the chromosphere), its magnetic field, and the influence of motions on that field. The photosphere is dominated by the convective motions, which are organized on all scales from 1 Mm to at least 30 Mm in convective cells, where the sizes of the convective cells are inversely proportional to the velocity of the plasma they contain, and their lifetimes are proportional to the square of their size, with the smallest granules having a lifetime of roughly 5 minutes, while the largest have lifetimes of more than a day. The velocity pattern is generated by a superposition of granules of different sizes, and it sweeps magnetic flux into concentrated collections of all sizes from large sunspots, all the way down to the smallest concentrations with scales of less than 100 km. As discussed in the Introduction, the magnetic field undergoes a major change through the chromosphere, with a transition from plasma dominated dynamics ($\beta > 1$) to magnetic field dominated dynamics ($\beta << 1$).	In addition, the small chromospheric pressure scale height implies a large gradient in the sound speed which creates an almost impassable barrier for all magnetosonic (compressional) waves generated in the photosphere. Such waves turn into shock fronts, are reflected by the large wave-speed gradient, and can deliver a substantial amount of energy to the chromospheric plasma. 

Other complications that arise are (i) that radiation of all wavelengths becomes decoupled from the plasma and can leave the Sun without being further absorbed or scattered. The decoupling is dependent on wavelength and so the details of the radiative transfer are important when trying to understand the energetics of the chromosphere and (ii) that the plasma changes from being very weakly ionised at the photosphere to fully ionised at the base of the corona. This has significant implications for both transport and dissipation of currents. These disparities in fundamental physical properties over such a small distance, as well as the intrinsic time-dependence, make any sort of modelling very difficult.
 
\begin{figure}
\centering
\hbox{
\includegraphics[height=8cm]{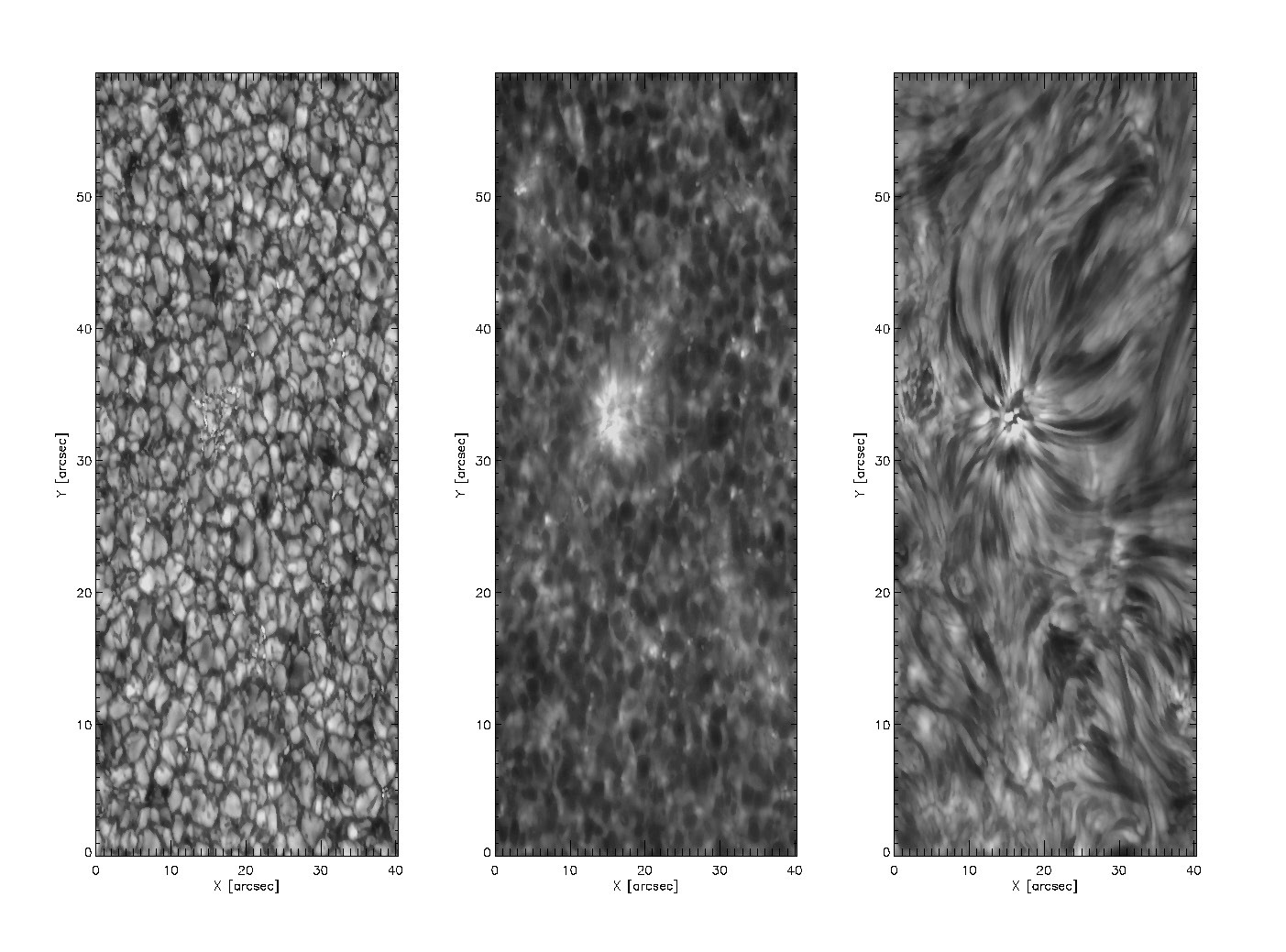}
}
\caption{From left to right are observations of the quiet solar atmosphere in white light, Ca II H and $H_\alpha$. The axes are in units of arcsec, corresponding to roughly 725 km on the sun. The observations are made with the Swedish 1-meter solar telescope at La Palma, by Luc Rouppe van der Voort on June 18., 2006. }\label{fig:sst_boris}
\end{figure}

When defined as a temperature interval, the chromosphere is split into two different parts. The lower is dominated by waves generated by the photospheric granules. Because the wave speed has such a large gradient, the waves shock in the lower chromosphere creating a pattern of colliding shock fronts. That pattern is prominent in the quiet sun where the magnetic field is generally weak and can be seen in the central panel of Figure~\ref{fig:sst_boris} taken with a filter centered on the Ca II H line. The same region observed in the $H_\alpha$ line, appears dramatically different (see the right panel of Figure~\ref{fig:sst_boris}), and since the prominent $H_\alpha$ line is the reason for naming the chromosphere, the part of the chromosphere showing the interacting shock fronts is now named the``Clapotisphere" \citep{2007ASPC..368...27R} by some. [For comparison a white light image is shown in the left panel of Figure~\ref{fig:sst_boris}.] An observation of the solar chromosphere in $H_\alpha$ is dominated by the structure of the magnetic field. Now the chromosphere looks like the fur of a haired animal, with dark and bright hairs with a width down to the resolution limit of our present instruments. The thin structures are due to the magnetic field's ability to isolate the plasma across the field direction making each strand a separate one-dimensional atmosphere. This very different look of the chromosphere is due to the change in the plasma $\beta$, so that the lower chromosphere is dominated by the plasma, while the upper region where $H_\alpha$ is produced is dominated by the magnetic field. 

\subsection{Energetics}
The energy balance in the solar chromosphere is quite subtle. Thermal conduction plays little role for two reasons. Since most of the chromosphere is at a temperature greater than the photosphere, there is no heat conduction into the chromosphere from below. On the other hand the corona has a temperature in excess of 1 MK, and energy is conducted downwards to the cooler regions (Section 3). The heat flux is strongly temperature-dependent: $F_c = \kappa_0 T^{5/2} \nabla T$ where $\kappa_0 = 10^{-6}$ ergs $\rm cm^{-1} s^{-1} K^{-7/2}$. One can estimate the average $F_c$ in the chromosphere by simply taking an average scale for the temperature gradients of 1 Mm and an average temperature of $10^4$ K, which yields a heat flux $F_c$ of order 1 erg cm$^{-2}$ s$^{-1}$, a completely negligible value. [The importance of this downward heat flux for coronal structure is discussed in Section 3.] The large outward radiative flux from the photosphere could also potentially heat the chromosphere, but the chromosphere is generally optically thin and does not absorb a significant part of the incident radiative spectrum. In fact radiation is the main coolant of the chromosphere, so there is no way radiation can sustain its temperature.

The two remaining possibilities are the dissipation of shock waves initiated in the solar photosphere, and the dissipation of magnetic energy. The magnetic energy arises because the photospheric mass motions inevitably lead to magnetic perturbations, for example in Alfven or magnetosonic waves. Both of these mechanisms transport energy from the photospheric kinetic energy reservoir, which can be estimated to be roughly 1000 ergs cm$^{-3}$ by assuming a mass density of $2 \times 10^{-7}$ g cm$^{-3}$ and a typical velocity of 1 km/s. The velocities can deliver an amount of work over a granular length scale and over a granular lifetime of $3 \times 10^8$ ergs cm $^{-2}$ s$^{-1}$. The kinetic energy available in the photosphere is then able deliver more than enough energy to heat the chromosphere (estimated to be $3 - 14 \times 10^6$ ergs cm$^{-2}$ s$^{-1}$, see references in \cite{2006ApJ...646..579F}) and the corona (estimated to be roughly $10^6$ ergs cm$^{-2}$ s$^{-1}$, at least for the quiet Sun). However, it turns out that the sonic portion of the shock waves does not contain enough energy to maintain the chromosphere's temperature \citep{2005Natur.435..919F,2006ApJ...646..579F,2007PASJ...59S.663C} even in the quiet sun. This leaves the options of heating either via pure Alfv\'en waves, the magnetic part of magnetosonic waves, by the reconnection of oppositely directed magnetic field components, or by other forms of current dissipation.

The injection of Poynting flux (${\bf S}$) into the chromosphere and corona happens primarily in the photosphere.  For a perfectly conducting plasma, one can write this as:
\begin{equation}
{\bf S} = B^2{\bf V}/4\pi - (\bf{V \cdot B}){\bf B}/4\pi
\end{equation}
where ${\bf V}$ and ${\bf B}$ are the velocity and magnetic field vectors. Breaking these down into components in the plane of the photosphere and in the vertical direction, the first term corresponds to direct injection of magnetic flux into the corona. Such emerging flux is recognised as a key component of coronal activity, in particular large flares (Section 4). The second term is commonly discussed in terms of fast and slow photospheric motions. For present purposes, assume that the typical magnetic concentration in the photosphere has a magnetic field strength of 1 kG  and that the typical velocity is roughly 1 km/s. Then, for motions in the plane of the photosphere, and no emerging flux, one finds $S_z = \frac{1}{8 \pi} |B|^2 u_t sin(2 \phi)$ where $u_t$ is the horizontal component of the velocity antiparallel to the horizontal magnetic field component. This is the stressing of the field by the horizontal motions, and will be discussed in greater detail in Section 3. The maximum value of $S_z$ is roughly $4 \times 10^8$ ergs cm$^{-2}$ s$^{-1}$. Of course the filling factor for the magnetic field is quite small but the amount of energy available where there is a magnetic field is very large. 

At some point the magnetic field will be stressed sufficiently that the associated free magnetic energy will be converted into heat either through the dissipation of electric currents, through magnetic reconnection events or through dissipation of waves and shocks.  As it evolves, the field will seek a minimum energy state, and for low-$\beta$ plasma this will be a potential field, or perhaps a force-free field characterized by field-aligned electric currents. Note that the most readily accessible minimum energy state need not be potential \citep{2010A&A...521A..70B}. The force-free state is present in the low corona, but not necessarily in the plasma-dominated lower chromosphere. However, if we assume that the chromosphere also has a force-free magnetic field, and the magnetic resistivity is constant, then the magnetic energy that may be converted into heat simply scales with the available magnetic energy, i.e. as $B^2$. All else being equal, the release of energy should decrease with height, and so the heating should be large in the chromosphere on the basis of this argument alone. The theoretical dissipation scale in the outer solar atmosphere is on the order of a few meters based on either resistive diffusion or collisionless processes and a typical temperature of $10^6$ K and density of $10^9$ $cm^{-3}$, so it would appear that the dissipation happens in a very small volume with high velocities as a consequence. However, the topological change occuring in reconnection also facilitates energy release over much larger scales through relaxation of stressed fields, particle acceleration, shocks etc. \citep{1953JWD, 1964NASSP..50..425P}. The dissipation of magnetic energy as the main heating mechanism in the chromosphere means that there are relevant scales from just a few tens of meters to tens of megameters. If the dissipation happens in the same way in the chromosphere as it does in the corona, it happens in a bursty fashion, so the relevant time scales are seconds for the individual dissipation events, to tens of hours for the stressing of the magnetic field on large scales. 

\subsection{Modern analysis of the chromosphere}

Major strides forward have been taken in recent years using space-based observations and MHD modelling. Given the short spatial and temporal scales, observations with short exposure times are needed, as well as high spectral resolution from everywhere in the 2D field of view with good spatial resolution. These are challenging requirements. The Hinode and Interface Region Imaging Spectrograph (IRIS) missions were both launched with the aim of understanding the outer solar atmosphere: implications for the corona from Hinode results are discussed in Section 3. Hinode has imagers with passbands at the wavelengths of the strong chromospheric spectral lines, so as to understand the linkage between corona, TR and chromosphere. IRIS is focused on the chromosphere, with instruments designed to understand the chromosphere and lower transition region. 

One of the discoveries of Hinode has been the very long thin spicule-like structures at chromospheric temperatures that penetrate many megameters into the hot corona, of which at least a subgroup seem to be ballistically driven by shock waves, while another subgroup seems to be longer lived and disappear while they still penetrate far into the corona (Type 2 spicules: \cite{2007Sci...318.1574D}). We have for some time known that the dynamics of the chromosphere and transition region produce correlations between the intensities of spectral lines formed in the chromosphere, while there is much less, if any, correlation between the intensity of spectral lines formed in the chromosphere and the transition region. We also know that there seems to be a correlation between the non-thermal width of spectral lines and their intensity. These correlations can only be reproduced by models that are able to catch the dynamics on all scales, and so far that has not been done. 

As we have made clear, modelling the chromosphere is challenging, but is now being undertaken successfully by some groups \citep{2012ApJ...753..161M}. The chromosphere cannot be modeled analytically, so multi-dimensional numerical 3-D MHD models are needed. Even then there are problems converting model (or simulated) time- and length-scales to those inferred from the observations.  Such models need to include radiative transfer and scattering and require a resolution which is high enough that the dynamical scales can be resolved. Further physics has been modeled such as non-equilibrium hydrogen ionization \citep{2007A&A...473..625L} and the effect of a large fraction of neutral atoms in the atmosphere, resulting in a generalized Ohm's law \citep{2012ApJ...753..161M}, both of which have an effect on chromospheric energetics. However not all of the physics can be included in the same numerical simulation. Even though the models have led to new discoveries and a deeper understanding of the solar chromosphere, it is questionable whether the simulations actually capture all of the small scale dynamics present. 

The chromosphere is hiding the answers to a number of important questions. The most important might be how the mass transfer from the photosphere to the corona happens, but another and most likely connected question is what type of magnetic heating happens in the chromosphere, the transition region and corona: are they the same and will we be able to identify the initiation of large flares by observing the chromosphere? The last major question is what produces the so called first ionization potential (FIP) effect \citep{2004ApJ...614.1063L} in which the observed abundances of metals with low first ionization potential seems to be higher in the corona than in the photosphere, and a satisfactory physical explanation for that has not yet been put forward. The obvious conclusion is that it is a electromagnetic effect, but how exactly and why relatively more metals with low ionization potential are present in the corona is unknown.

\section{The non-flaring closed corona}

The general properties of the non-flaring, closed corona are well known. The basic structures are magnetic loops that connect regions of opposite surface magnetic polarity. Loops can be considered as mini-atmospheres with plasma properties determined by the energy input. \cite{2010LRSP....7....5R} provides a useful summary. [It should however be stressed that when one sees, for example, a coronal loop in EUV emission, one is looking at a collection of magnetic field lines that happen to be illuminated, not at an isolated bundle of field in a surrounding plasma.] AR loops have typical scales of 25 - 120 Mm, the temperature at the peak of the emission is around $10^{6.6}$ K \citep[e.g.][]{2012ApJ...759..141W}, and they are bright because they are relatively dense (a few $10^9$ cm$^{-3}$). The quiet Sun is cooler and less bright and exhibits larger structures in excess of 100 Mm such as streamers (Section 5) and interconnecting loops.

A major difficulty in understanding the corona is the determination of its magnetic field. In general, the Zeeman effect is not detectable due to the thermal broadening of EUV emission lines. An exception are coronagraph observations of the outer corona using Fe lines in the visible or IR \citep[e.g.][]{2000ApJ...541L..83L,2004ApJ...613L.177L}. It is to be hoped that the Daniel K Inouye Solar Telescope (DKIST) will permit improved observations. Though not done regularly, the field strength can be constrained by microwave maps of the primarily gyrosynchrotron radiation, together with optically thin EUV temperature and density diagnostics \citep[e.g.][]{2002ApJ...574..453B}. Future microwave arrays such as the Frequency Agile Solar Radiotelescope (FASR \cite{2004P&SS...52.1381B}) and its pathfinder, the Extended Owens Valley Solar Array (EOVSA) will do this with high spatial, temporal and spectral resolution, and in principle the variations of magnetic energy can be deduced, although strong (few hundred G) fields are required. 

Theoretical concepts can also be used to deduce coronal magnetic field properties. One approach is coronal seismology which relies on conjectured properties of oscillating coronal magnetic loops \citep[e.g.][]{1999Sci...285..862N,2005LRSP....2....3N,2007Sci...317.1192T}. Given a period of loop oscillation, a loop length and an estimate of the density, the magnetic field magnitude can be obtained. Typical values are 10 - 20 G. Force-free reconstruction of the coronal field from photospheric measurements \citep[e.g.][]{2006SoPh..235..161S,2008SoPh..247..269M} are widely used. This is a difficult task, beset with observational and computational difficulties, and solutions are non-unique \citep{2009ApJ...696.1780D}. More reliable extrapolation results will be possible when the chromospheric vector magnetic field is measured. However, unlike seismological methods, a 3D field can be constructed. Also useful are EUV and X-ray images and movies, although they provides no information on field magnitude. Since the field and plasma are frozen, and strong thermal conduction  leads to nearly-isothermal conditions in the high corona, plasma structures are very often taken to be an indicator of field geometry. This in turn provides information on both large-scale structures such as separatrices (Section 4) and the small-scale structure present within a large-scale field: for example \cite{2012ApJ...755L..33B} identified distinct loops with scales of 1". Results from the Hi-C rocket flight suggests that field topology on sub-arc second scales can be inferred from images and movies \citep{2013Natur.493..501C}.

\subsection{Energy requirements and atmospheric thermal structure}

The energy requirements of the various coronal regions date to \cite{1977ARA&A..15..363W} and are $10^7$, $10^6$ and a few $10^5$ ergs cm$^{-2}$ s$^{-1}$ for ARs, coronal holes and quiet Sun respectively. The Poynting flux at the base of the chromosphere is described in Section 2. Of the hydromagnetic waves generated by the small-scale motions in the photosphere, only the Alfv\'en wave can reach the corona, and these have an upward Poynting flux of $\rho V_A\delta V^2$ where $\rho$ is the mass density, $V_A$ the Alfven speed and $\delta V$ the wave amplitude. For slow (e.g. granular) motions ($< 1$ km/s) the Poynting flux is $VB_t B_a/4\pi$, where $B_a$ and $B_t$ are magnetic field components perpendicular to, and in the plane of the photosphere respectively. If one takes typical values: $B_a$ = 150G, $B_t$ = 50 G, V = 1 km/s, $\delta V$ = 30 km/s, the Poynting flux for both Alfv\'en waves and slow injection is a few $10^7$ ergs cm$^{-2}$ s$^{-1}$, so in principle sufficient energy is injected.  

How a corona forms as a consequence of heating is well understood, and involves the small but important transition region (TR) between corona and chromosphere. The behaviour of the corona/TR system is governed by three energy transfer processes: thermal conduction from the corona to the lower atmosphere, optically thin radiation to space and mass flows between the upper chromosphere, TR and corona. \cite{1979ApJ...233..987V} noted that the TR can be defined as the region above the top of the chromosphere and below the location where thermal conduction changes from being an energy loss to an energy gain. In a static loop, the TR structure is determined roughly by a balance between a downward heat flux from the corona and radiation. Indeed the radiation from the TR is of order twice that from the corona \citep{2012ApJ...752..161C}, despite its small size ($< 10\%$ of a typical loop). 

We can use this to understand how a corona forms. Consider a low density coronal loop that is heated rapidly. Heat is conducted from the corona, but the TR is not dense enough to radiate it away. The only way for the TR and chromosphere to respond is by an enthalpy flux into the corona, increasing the density until a steady state is reached \citep[e.g.][]{1978ApJ...220.1137A}. Turning the heating off has the opposite effect. As the corona cools, the conduction is not strong enough to power the TR radiation. Instead an enthalpy flux is set up to power the TR and the corona drains \citep[e.g.][]{2010ApJ...717..163B}. A simple physically motivated account of these processes is in \cite{2008ApJ...682.1351K} and \cite{2012ApJ...752..161C}.

It is very important to understand that the TR is not at a fixed location or temperature: these adapt in response to the conditions within any loop. Thus for a hot AR loop, Fe IX emission is from the TR, as is seen with the moss. For a cooler loop (1 MK peak temperature say), plasma only below 0.5 MK is TR. Misconceptions of the TR persist in the literature with reference to TR plasma or TR lines or in the TR. Such definitions are meaningless without understanding the overall context of the loop for which they are made.

\subsection{Magnetic field equilibrium and stability}

It is, in principle, straightforward to develop MHD models of individual coronal loops and the more general large-scale coronal structures such as streamers, arcades, prominences etc, and examples can be found in Sections 4 and 5. With any such magnetic field, two questions are important: (i) does an equilibrium exist for given boundary conditions and (ii) if it does, is it linearly (and non-linearly) stable? Seen in movies, the corona is in continual motion, but large eruptive flares are rare, so most of the time the magnetic field configuration is in, or close to, a state of equilibrium, or undergoing a weak instability that does not destroy the global field configuration. So perhaps the real question is not: why do large flares occur? Rather it is: why do they not occur more often? It is in fact quite difficult to eject material from the Sun (Section 4).  

One reason for this stability lies in the line-tying of the magnetic field lines at the photosphere. The high density there means that magnetic field lines cannot move in response to a perturbation originating in the corona. For example, in a cylindrical arcade extending over $\pi$ in the corona, the most unstable ($m = 1$) mode cannot arise, and the other modes are absolutely stable over a wide range of equilibria \citep{1983SoPh...87..279H,1986ApJ...309..402C}. For a cylindrical loop tied at both ends, the threshold for the (destructive) kink instability increases by of order 50\% over that expected in the laboratory \citep[e.g.][]{1979SoPh...64..303H}. In both these geometries global resistive instabilities such as the tearing mode are suppressed except for the most eccentric field conditions. Also, proof of linear instability does not mean that the pre-instability field structure is entirely destroyed since the instability can saturate (non-linearly) at a low level. This is important because the vast majority of flares are not associated with a CME, so require very significant energy release but not the destruction of the field configuration.

\subsection{Coronal heating: MHD aspects}

Coronal heating has been divided for many decades into studies of wave heating (Alfv\'en waves generated at the photosphere) and heating by small-scale coronal reconnection. Alfv\'en wave heating has been discussed extensively \citep{2005LRSP....2....3N,2006SoPh..234...41K} and requires structuring in the atmosphere and/or magnetic field to get dissipation by either phase mixing \citep[e.g.][]{1984A&A...131..283B} or resonance absorption \citep[e.g.][]{1978ApJ...226..650I}. We will not discuss this further but note that \cite{1998ApJ...493..474O} pointed out that resonance absorption became an impulsive heating process when feedback from the chromosphere was included. A different approach to wave heating is due to \cite{2011ApJ...736....3V}, \cite{2012ApJ...746...81A} and \cite{2013ApJ...773..111A}. They argued that the generation of Alfv\'en waves at two loop footpoints, with different wave properties at each, would lead to a turbulent cascade as the counter-propagating waves interacted, eventually reaching dissipation scales. The heating is highly time and spatially dependent, with bursts of energy being released on top of a low background level of dissipation.

The consequences for the coronal magnetic field of slow photospheric motions have been studied extensively. For AR heating, the conditions imposed by this scenario are quite severe \citep{1988ApJ...330..474P}, with a ratio $B_t/B_a$ of order 0.25 - 0.4 being required. Here $B_t$ is a typical field stength in a direction between the loop footpoints and $B_a$ the field in the direction parallel to the photosphere, where we assumed that a curved loop has been straightened out. This implies that the coronal field must resist any desire to dissipate before a stressed condition is reached. The energy released if the dissipation returns $B_t$ to zero is of order $10^{23} - 10^{26}$ ergs, which led to the term nanoflares. It takes in excess of $10^4$ secs to build up the energy in a nanoflare, with implications discussed shortly. Thus, in this scenario, the corona is maintained at its temperature by a swarm of nanoflares, each occurring in a small volume \citep{1988ApJ...330..474P,1994ApJ...422..381C}. Theoretical evidence for nanoflare heating with such conditions on $B_t/B_a$ is not yet convincing. \cite{2005ApJ...622.1191D,2009ApJ...704.1059D} and \cite{2013A&A...560A..89B} have shown that large values of shear are possible, but computational limitations are a concern in transferring their results to the real corona. 

There is a burgeoning body of work that treats the corona as a global system in an MHD simulation, including a chromosphere, and attempts to impose realistic photospheric motions \citep[e.g.][]{2005ApJ...618.1020G,2013A&A...550A..30B}. While these models do produce something that looks like a corona, the temperatures are low (in part because a large enough simulation is unfeasible), heating occurs near the base of any loop, and is attributed to ohmic heating. However, it may be this ohmic heating may be an artefact of numerical resolution. The level of energy release at say an x-point is rather small: this was pointed out by Dungey as long ago as 1953 (see also \cite{2014PJC}). Magnetic reconnection is a facilitator of energy release: shocks, waves, particle acceleration, turbulence away from a reconnection site are far more important.

\begin{figure}
\begin{centering}
\hbox{
\includegraphics[height=4cm]{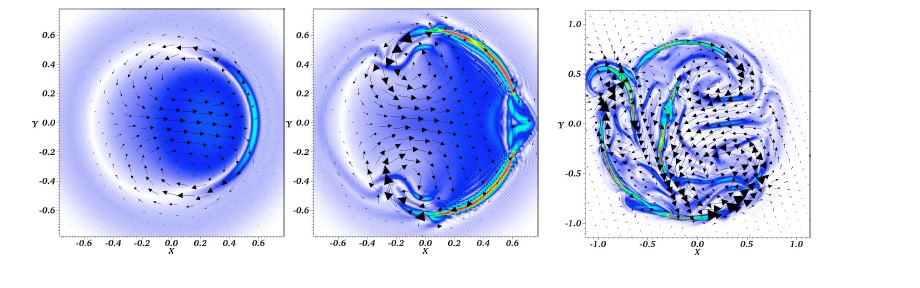}
}
\end{centering}
\caption{Evolution of a kink instability. The three plots show current magnitude (coloured background) and velocity vectors (arrows) at three different times. Note the formation of very fine-scale structure in the final plot. From \cite{2009A&A...506..913H}.}\label{fig:hood_browning}
\end{figure}

Addressing the real dissipation processes is difficult, and requires more local models which should be seen as complementary to the global ones. As an example, consider the non-linear evolution of the kink instability \citep{2008A&A...485..837B,2009A&A...506..913H}. The energy release here is more appropriate for a microflare or a nanoflare storm than AR heating, but it makes an important point. Figure~\ref{fig:hood_browning} shows the current and velocity at three stages of the instability. First (left panel), regions form where oppositely-directed field components are pushed together, initiating forced magnetic reconnection. A single current sheet evolves and then begins to fragment (2nd panel). Finally, a wide range of small-scale structures emerges (3rd panel) which then dissipate, leading to a magnetic field structure that is stable to the kink mode. 

While it is tempting to invoke the language of turbulence here, the numerical grid does not permit the construction of a meaningful distribution of scales. But what is clear is the evolution of a smooth magnetic field into a very fragmented one through the instability. An important aspect is that the main energy release is not at the obvious current sheet, but may be due to slow shocks driven by the vortical flow, with secondary heating in the rest of the turbulent structure \citep{2013SPD....4420002B}. These dissipating structures need to be resolved properly to actually understand what is going on. So calling everything `ohmic heating' is a little lazy and possibly misleading.

Another common approach is use the reduced MHD approximation which assumes $B_t \ll B_a$. A number of papers \cite{2008ApJ...677.1348R,2010ApJ...722...65R,2012A&A...544L..20D} have demonstrated that dissipation occurs readily for a range of photospheric motions with discrete heating events occurring. However, it is impossible for RMHD models to account for AR heating. The required value of $B_t/B_a$ is beyond the viability of the model. Indeed, one could say that dissipation occurs too easily in these models: this may be due to assumptions about how the coronal field responds to photospheric motions. For example, adjustments to neighbouring equilibrium states are forbidden, so that dynamic evolution must occur instead.

\subsection{Deducing properties of heating mechanisms}

In the absence of direct observations of coronal magnetic field dynamics, inference of heating processes relies on the interpretation of images and spectra. In both cases, it is important to have comprehensive temperature coverage and data from a wide range of emission lines can provide information on plasma parameters such as mass motions, density etc. using spectral techniques. The results have been mixed. Density-sensitive line pairs (i.e. same element, same ionisation state, same temperature, different transition) can provide an absolute measurement of the electron density. Combined with an estimate of the density from an (assumed unresolved) image, this provides a handle on the filling factor: the ratio of the volume radiating to the actual volume. In turn, this can provide information on the fundamental scales associated with coronal heating. \cite{1997ApJ...478..799C} obtained scales of $<$ 100 km. Future high-quality spectroscopic data has the potential to provide such information.

Analysis of coronal mass motions provides another possible diagnostic. An important aspect of magnetic reconnection, at least in a simple form, is the prediction of high-speed reconnection jets which have been detected at magnetopause and solar wind reconnection sites. In the corona, these should lead to Doppler shifts or non-thermal broadening (if multiple sites are convolved: \cite[e.g.][]{1996SoPh..167..267C}) well in excess of 100 km/s. While individual jets have been well observed over many years (see, for example, \cite{1997Natur.386..811I,1998ESASP.421..233K}), and more recently coronal manifestations of spicules \citep{2009ApJ...707..524M,2010A&A...521A..51P}, line broadening in active regions of the magnitude predicted is not seen \citep{2012ApJ...754..153D}. Reasons may include: the temperature of the initial jet is not measured, the jet interacts with the surrounding magnetic field and thermalizes rapidly or the initial jet density is small (as would be the case in a low filling factor corona), with negligible emission measure \citep{1993SoPh..148...43Z,1998ESASP.421..233K} or a nanoflare may be too small \citep[e.g.][]{2013ApJ...770L...1T}.
  
\begin{figure}
\begin{centering}
\hbox{
\includegraphics[width=0.45\textwidth]{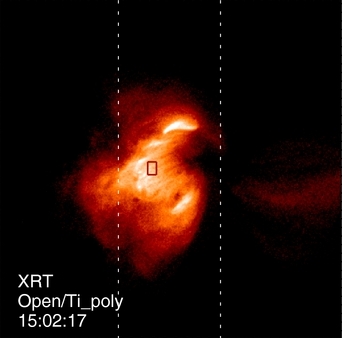}
\includegraphics[width=0.45\textwidth]{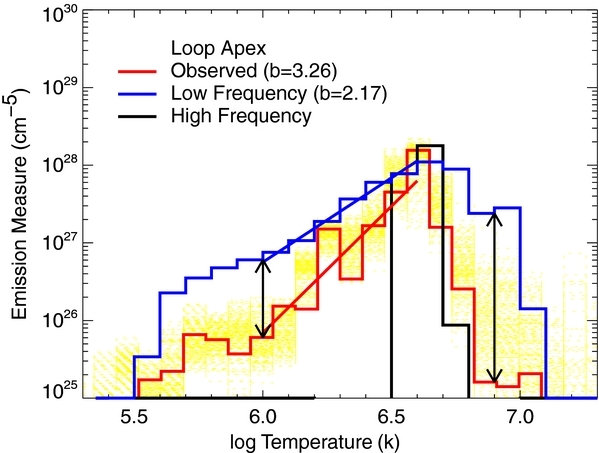}
}
\end{centering}
\caption{On the left: AR loops seen on the disk by the Hinode X-ray telescope (XRT). The small rectangle in the centre corresponds to loops in the AR core that were analysed by the Hinode EIS instrument (from \cite{2011ApJ...734...90W}). On the right is the emission measure as a function of temperature from EIS. The red line is the observed value, and the blue and black correspond to low and high frequency nanoflares as defined by \cite{2011ApJ...734...90W}.}
\label{fig:warren}
\end{figure}

\begin{figure}
\begin{centering}
\hbox{
\includegraphics[height=4cm]{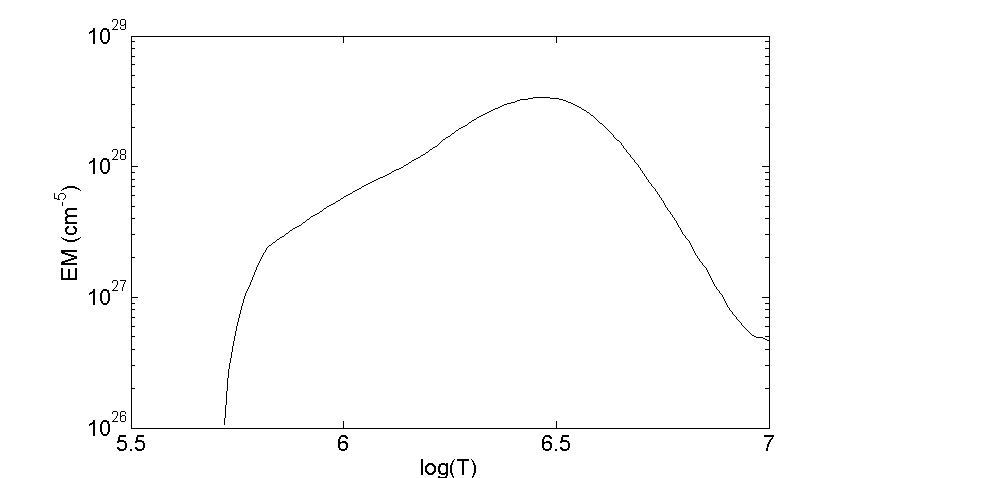}
}
\end{centering}
\caption{The EM-T profile from a single nanoflare obtained from the EBTEL model \citep{2012ApJ...752..161C}.}\label{fig:em_T}
\end{figure}

At the present time, the most promising approach by which progress is being made in understanding coronal heating is through emission measure analysis that evaluates the dependence of the emission on temperature. Figure~\ref{fig:warren} shows an image of an active region core \citep{2011ApJ...734...90W} and other AR loops have been analysed by a number of workers \citep{2011ApJ...734...90W,2012ApJ...759..141W,2011ApJ...740..111T,2012ApJ...756..126S}. Using many spectral lines from the Hinode EIS instrument, as well as data from SDO and Hinode, they were able to construct the EM(T) profile over a wide temperature range. One popular heating model, that due to relatively infrequent nanoflares that involve slow injection of energy followed by rapid dissipation, predicts that $EM(T) \sim T^2$ as shown in Figure~\ref{fig:em_T} \citep{1994ApJ...422..381C,2004ApJ...605..911C}. \cite{2011ApJ...734...90W} found $EM \sim T^{3.1}$ and concluded that infrequent nanoflares were not heating the corona, instead that steady heating was responsible. Subsequent parameter studies showed that there were some ARs that did fit the original prediction, but that many did not. 

The real question here is: what is the frequency of the nanoflares? The key parameter is the ratio of the time taken for a loop to cool from its initial heated state to below 1 MK, to the recurrence rate of the nanoflare along a heated sub-loop. The former is of order 1000 - 3000 secs, scales with the loop length \citep{1995ApJ...439.1034C}, and is almost entirely independent of the nanoflare energy \citep{2014ApJ...784...49C}. Recently it has been shown that these AR observations cannot be reproduced if the nanoflare recurrence time is much under 500 s, and it needs to lie in the range 500 - 2000 s. In addition, the nanoflare energy distribution should take on the form of a power law, with the delay time between each nanoflare proportional to the energy of the second nanoflare. If all these conditions are met, then the AR results can be accounted for \citep{2014ApJ...784...49C}.

This result poses questions for coronal heating by nanoflares. Namely, it is impossible to power the corona if the nanoflare involves the complete relaxation of the field to a near-potential state since, as was noted above, the time to rebuild the nanoflare energy is long ($> 10^4$ s). Instead, nanoflares must involve a small relaxation of the field to a slightly less stressed state, which then permits a rapid rebuilding. How this happens in the framework of MHD is unclear.

A second important clue from contemporary data concerns hot coronal plasma. Any impulsive heating process leads to a plasma component that lies well above the peak of the emission measure, of order $10^{6.7} - 10^{6.9}$ K, as seen on the right hand side of Figure~\ref{fig:em_T}. Detection of this would be a powerful result. This has been accomplished by several workers at this time \citep{2009ApJ...698..756R,2011ApJ...728...30T,2012ApJ...750L..10T}. Unfortunately characterising the physical properties of such plasma is very difficult. Low emission measures, ionisation non-equilibrium, the difficult nature of thermal conduction and presence of non-Maxwellian distributions make this an ongoing challenge.

\subsection{Discussion}
The last few years have seen remarkable progress in the understanding of the properties of the hot non-flaring corona. This has come about through the extensive data bases of the Hinode and SDO missions, modelling, both hydrodynamic and MHD, and the ability to forward model observables from these models. Yet major puzzles remain. The time-dependence of coronal heating mechanisms remains unclear, and associated with this is a lack of clarity of the plasma environment in which heating occurs (tenuous $< 10^8 ~{\rm cm^{-3}}$ or dense: a few $10^9 ~{\rm cm^{-3}}$). This in turn is likely to feed into the extent of any hot plasma component discussed above and/or the presence of accelerated particles in the non-flaring corona. The lack of detection of the latter in particular is a puzzle since magnetic reconnection is well known as a good facilitator of particle acceleration. 

Progress on these issues requires analysis of current data, modelling, and innovative uses of new data sources such as the Interface Region Imaging Spectrograph (IRIS) and the Nuclear Spectroscopic Telescope Array (NuSTAR) telescope. One viewpoint is that the smoking gun of coronal heating may lie in the presence or absence of these hot and energetic components. Unfortunately for the small nanoflare energies now being discussed, detection is likely to be even more difficult and may require long integrations over multiple sources to build up a real signal. An additional problem is that while nanoflares are often discussed as mini-flares, it seems most unlikely that the efficient acceleration occuring in flares that leads to 30\% of the energy going into accelerated particles persists in nanoflares. Much smaller fractions seem likely, enhancing the detection problem.

\section{Flares and eruptions}
\subsection{Introduction: flare and CME properties}
\label{intro}

\begin{figure}
\begin{centering}
\hbox{
\includegraphics[width=0.45\textwidth]{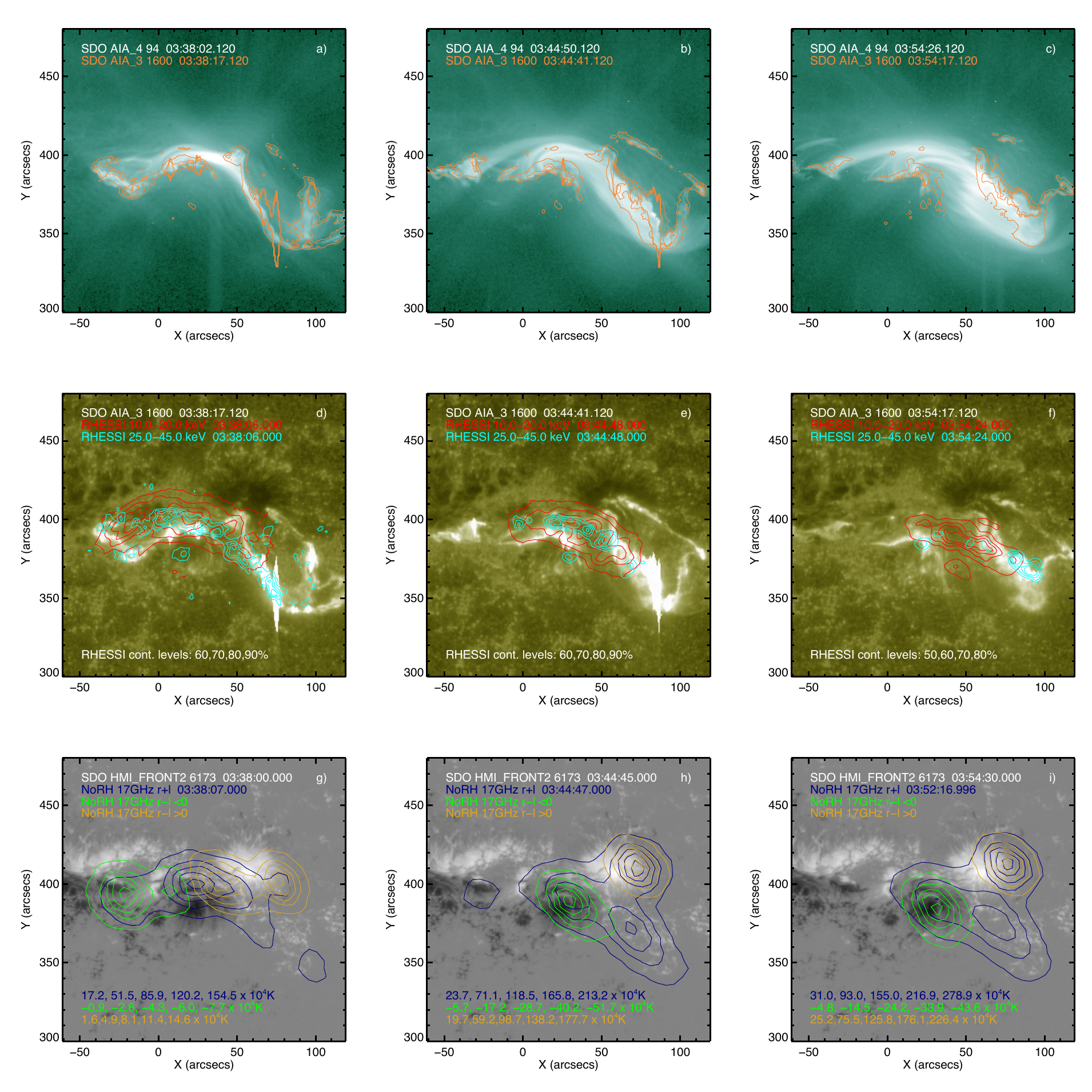}
\includegraphics[width=0.45\textwidth]{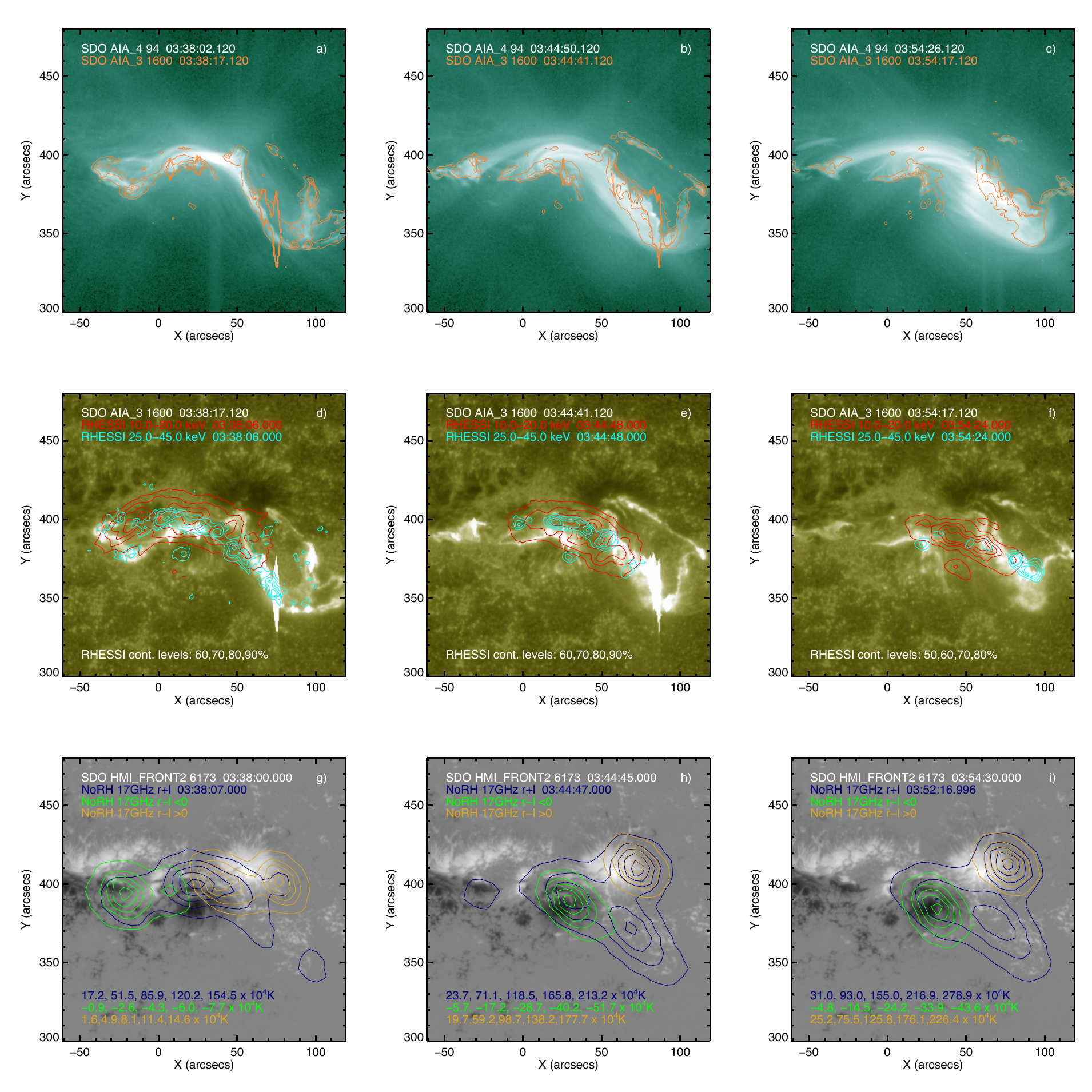}
}
\end{centering}
\caption{Two images from a flare on 9th March 2012. The LH panel shows UV emission from the chromosphere, dominated by a number of flare ribbons. Cyan intensity contours are hard X-ray footpoints, and red intensity contours are thermal emission with a strong coronal component. The RH panel shows hot coronal emission at around 10~MK structured into quite well-organised loops. Orange contours show the ribbon locations. From~\cite{2013arXiv1309.7090S}}\label{fig:niceflare}
\end{figure}

Solar flares are abrupt and dramatic outbursts of radiation in the solar atmosphere, almost always taking place in magnetic active regions, and spanning a range of scales in size and energy from the smallest (currently) observable events called `microflares' with energies of $10^{26}$ergs, to the largest great flares with energy of up to a few $\times 10^{32}$ ergs \citep[e.g.][]{2011SSRv..159..263H}. 
Flares involve the acceleration of large numbers of electrons and ions up to mildly relativistic energies, a minority of which have access to the interplanetary magnetic field. The remainder are contained in the closed magnetic structures of the lower solar atmosphere. Particle acceleration characterises the time of the main energy release in solar flares, and is a major focus of theoretical and observational attention. However, the flare is defined by its radiation burst, the majority of which is produced in the optical and UV parts of the spectrum \citep{2004GeoRL..3110802W} and emitted primarily by the solar chromosphere and photosphere in concentrated sources known as `footpoints' or `ribbons' depending on whether they have a point-like or a linear morphology. This emission is generated when the lower solar atmosphere is heated during a flare, and collisional heating by particles probably plays a considerable role in this, in addition to producing high energy radiations; hard X-rays (HXRs) from the electrons and $\gamma$-rays from the ions.  Other mechanisms, e.g. damping of Alfv\'en waves, may also be important in heating the lower atmosphere at depths where particles cannot readily penetrate \citep{2013ApJ...765...81R} The corona radiates mostly in extreme UV and soft X-rays during a flare, but the total energy is a small fraction of the bolometric radiated energy \citep{2012ApJ...759...71E}. The coronal emission is organised into beautiful, well-defined arcades, called flare loops. A composite image of a well-observed flare is shown in Figure~\ref{fig:niceflare}.

\begin{figure}
\centering
\hbox{
\includegraphics[width=0.45\textwidth]{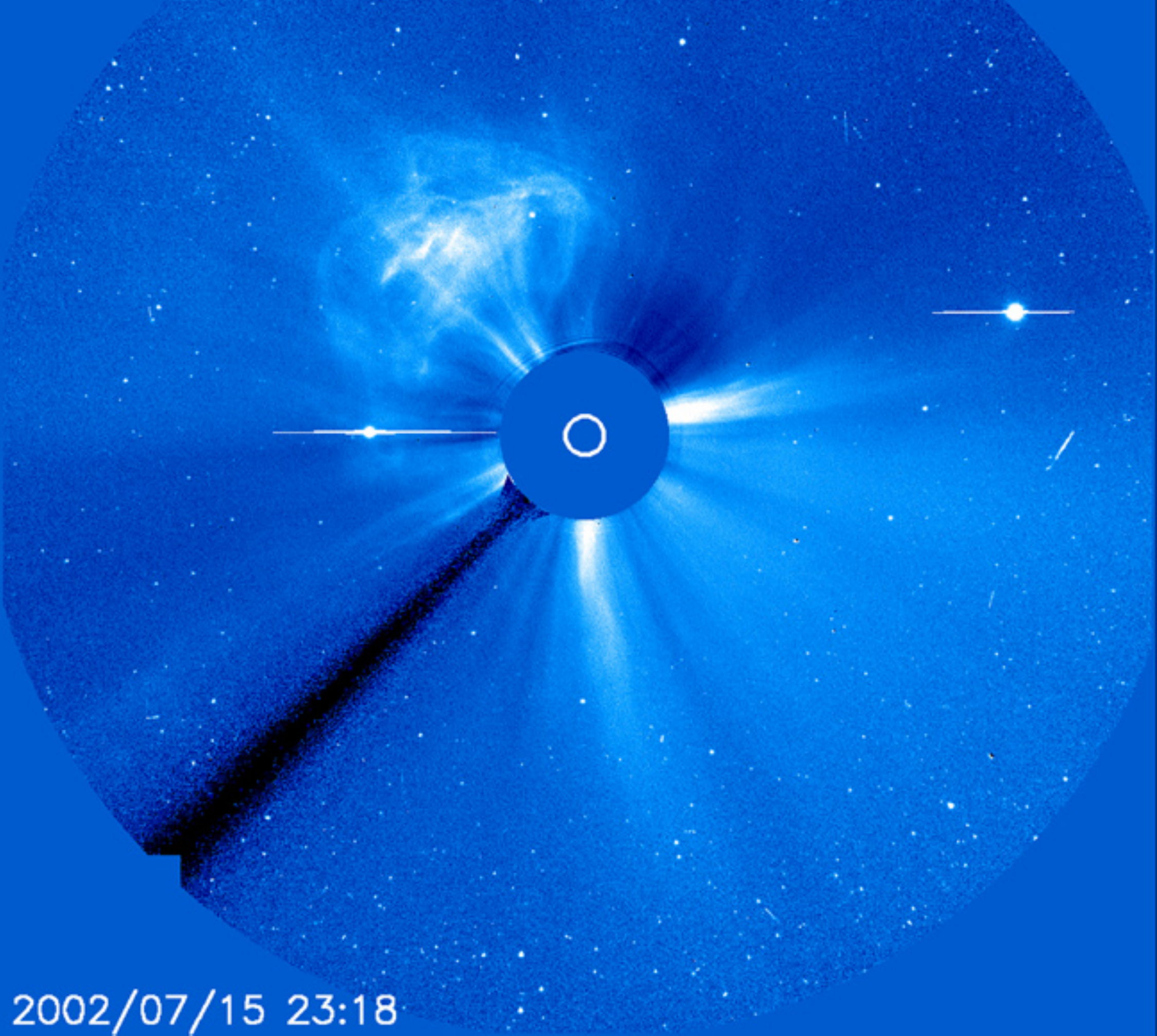}
\includegraphics[width=0.45\textwidth]{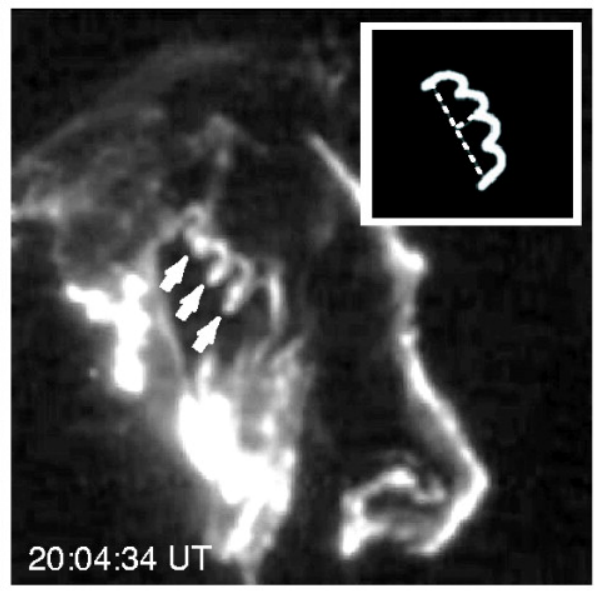}
}
\caption{A CME observed by the LASCO instrument on SOHO (left-hand panel), and a UV snapshot of the eruption during its related flare (right-hand panel), as reported by \cite{2004ApJ...611..545G}. The flare clearly shows an erupting twisted structure - with individual twists indicated by the arrows. Twist is also visible in the CME. Note, this is an extreme example of twist in an eruption; fewer turns are normally visible}\label{fig:niceCME}
\end{figure}

Flares take place in a variety of magnetic structures. Characteristic of large flares are active region filaments or filament channels, which are configurations including concave-upwards magnetic fields capable of supporting cool, dense plasma typically close to the solar surface and roughly parallel with the magnetic polarity inversion line. Smaller flares can take place in single or small groups of loops. The instability mechanisms are likely quite different in each. The magnetic reconfiguration that takes place in a flare can also lead to the ejection of magnetised plasma into the heliosphere, as a coronal mass ejection (CME) as shown in Figure~\ref{fig:niceCME}. CMEs typically travel around 500 $\rm{km\;s^{-1}}$ with a total mass (measured from Thomson scattering) of around $10^{15}$g. Statistically, more energetic flares are more likely to be associated with a CME \citep{2006ApJ...650L.143Y}. A CME's total energy 
has been estimated at half an order of magnitude greater than that radiated in a flare \citep{2012ApJ...759...71E} but a flare has a much higher energy density. The median time between the flare HXR peak and CME peak acceleration is around one minute \citep{2012ApJ...753...88B}. The overall picture is of a dramatic re-organisation of coronal magnetic structures which results in both upward-going and downward-going energy fluxes, with downward-going energy channeled by the (strong) magnetic field, resulting in heating and particle acceleration. A secondary response to both of these is expansion of chromospheric material into closed coronal field, usually termed `chromospheric evaporation' and visible clearly in spectroscopic observations. This results in the bright EUV-emitting flare loops. 

Large flares occur almost exclusively in solar active regions hosting sunspots; for example \cite{1970SoPh...13..401D} found that 7\% of large H$\alpha$ flares occur in regions with small or no spots, while 82\% of X1 flares were observed to occur in magnetically complex ``$\beta\gamma\delta$'' spots by \cite{2000ApJ...540..583S}. The fastest CMEs are associated with active region flares \citep{2005JGRA..11012S05Y,2012ApJ...755...44B}, but CMEs can also occur with the eruption of quiescent prominences outside active regions without an accompanying flare.
The magnetic field in an active region, which is readily observed emerging and developing at the solar photosphere, is the primary agent imposing the structure, and determining the evolution, of solar flares and CMEs at launch (the solar wind dynamical pressure and frictional stresses also plays a role in later CME evolution). The heart of the flare problem is to work out how energy stored in large-scale and organised coronal magnetic structures is released and transferred to scales at which it can be picked up by individual electrons and ions, resulting in heating and acceleration.

\subsection{Flare morphology, magnetic structure and magnetic topology}
There is a tendency for flares to occur close to the magnetic polarity inversion line in complicated, sunspot groups with a large amount of free magnetic energy. Flaring spot groups also exhibit rapid evolution in their strong-field regions: for example, \cite{2007ApJ...655L.117S} finds that the total unsigned magnetic flux within 15Mm of  the polarity inversion line increases by around 20\% in the 2 days prior to a major flare.
The bulk of the free energy is associated with elongated and twisted (i.e. current-carrying) magnetic fields - called flux ropes -  concentrated within a few 1,000~km of the magnetic polarity inversion line \citep[e.g.][]{2012ApJ...748...77S}. 
It is not yet possible to probe the current-carrying field in detail, but extrapolations of the magnetic field into the corona indicate that the bulk current-carrying structures shift in position during a flare, and change twist. While the field carrying most of the energy for the flare is organised into a relatively simple flux-rope structure on the scales of a thousand~km or so, there are doubtless smaller-scale structures embedded within this which may be critical for the flare triggering and evolution. It is estimated that at least 40\% of CMEs are flux-rope eruptions \citep{2013SoPh..284..179V}, the remainder appearing more like jets and outflows.

Though the energy for the flare appears to be stored in a substantial volume of the corona, it is focussed on release into a small number of compact patches, known as footpoints and ribbons. The footpoints are sites of the most energetic emission in optical and also in HXRs (which indicate high number densities of non-thermal electrons) and tend to have a point-like morphology. They are a subset of narrow, well-defined and elongated UV flare ribbons \citep{2006SoPh..234...79H}. 

\begin{figure}
\centering
\includegraphics[height=6cm]{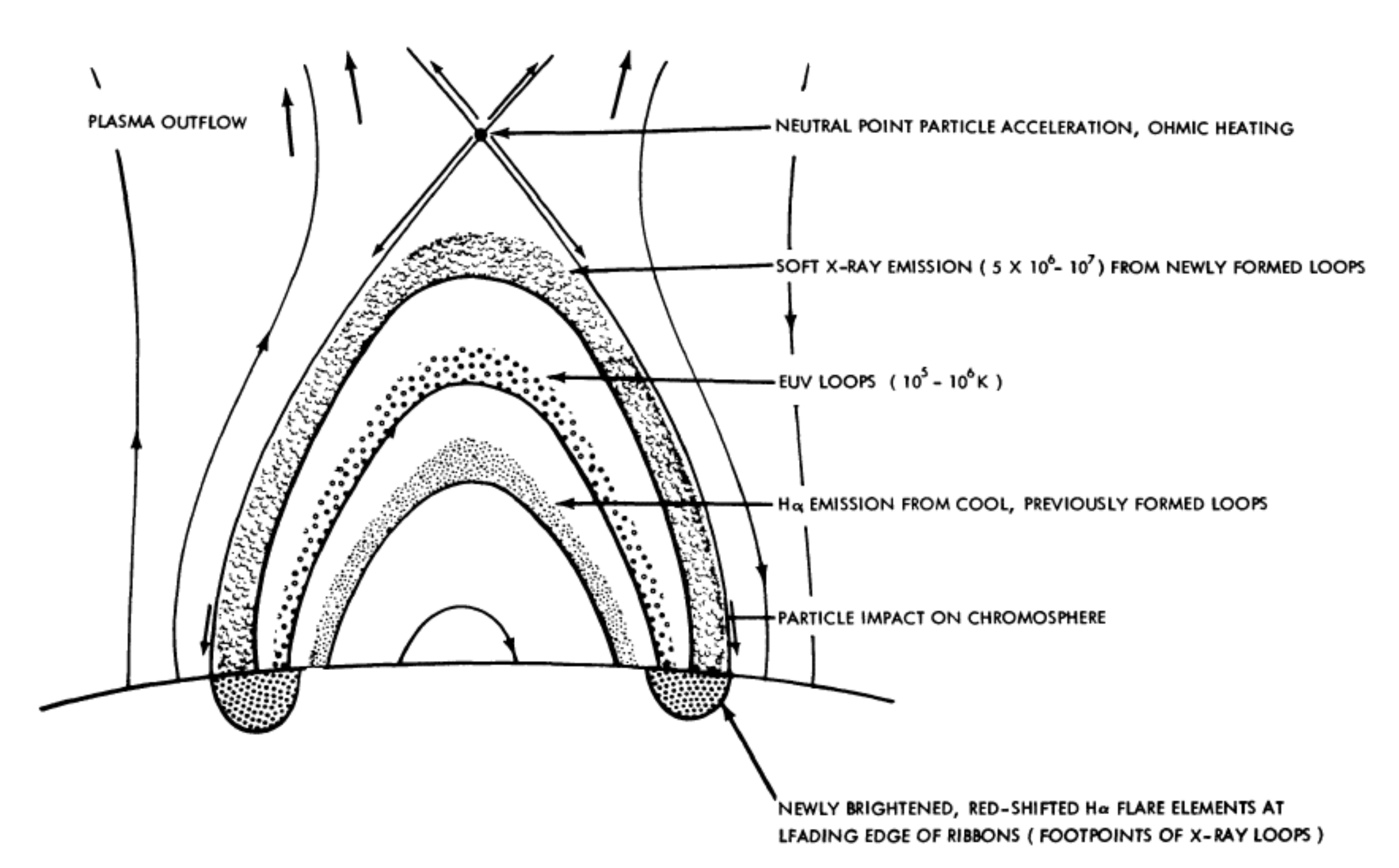}
\caption{A rendering of the 2-D CSHKP model by \cite{1980IAUS...91..217S} provides an illustration of the link between field topology and flare evolution. The field lines extending to the photosphere from the neutral point (X-point) are separatrices which, translated into the 3rd dimension, become separatrix surfaces. Energy is delivered to the lower atmosphere at or near these surfaces, and field advecting into the X-point from the outer regions results in the spreading of flare ribbons.} \label{fig:cshkp}
\end{figure}

The coronal magnetic field is rooted in a complicated photospheric field, and attempts to describe this call on notions of magnetic topology, which describe how one part of the field is linked to another.  A simple example of this applied to the late evolution of a solar flare is the so-called `CSHKP' model \citep[after the primary authors, ][]{1964NASSP..50..451C,1966Natur.211..695S, 1974SoPh...34..323H,1976SoPh...50...85K}, which provides a plausible explanation for the appearance and the slow spreading apart  of ribbons in the late phase of a flare in terms of the interface between post-reconnection closed loops and pre-reconnection unconnected field. Because it is a steady, two-dimensional view its applicability is restricted to the flare gradual phase, and it is used here to indicate why magnetic topology is of concern in understanding observed structures. In the CSHKP model (Figure~\ref{fig:cshkp}) reconnection between oppositely directed magnetic field occurs in the corona.  Reconnected field retracts from the reconnection region, both upwards and downwards, allowing free energy to be liberated from the larger-scale magnetic structure as the field relaxes to a lower energy state. The liberated energy, ducted along the magnetic field, heats the lower atmosphere near the ends of the just-reconnected field on opposite sides of the polarity inversion line. The next set of field lines brought into the reconnection region are rooted somewhat further from the polarity inversion, so that the location of heating moves outwards. Ribbons would arise from the extension of this 2-D model in an invariant direction, with their narrowness reflecting both the ducting of the energy along the field, and the rapid cooling of the lower atmosphere by radiation. 

\begin{figure}
\begin{center}
\hbox{
\includegraphics[height=4cm, width=0.45\textwidth]{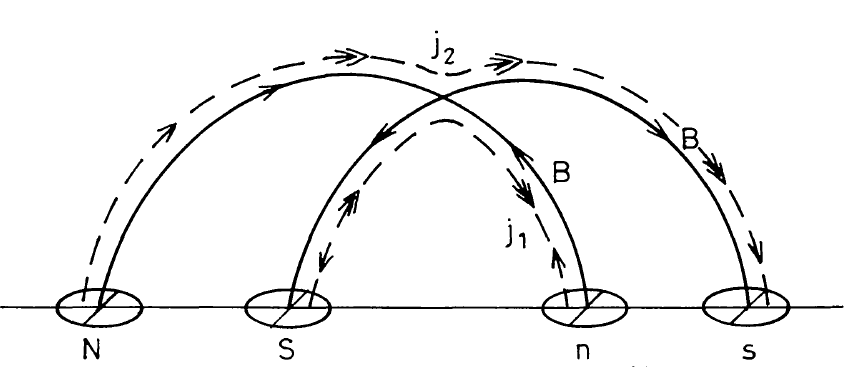}
\includegraphics[height=4cm]{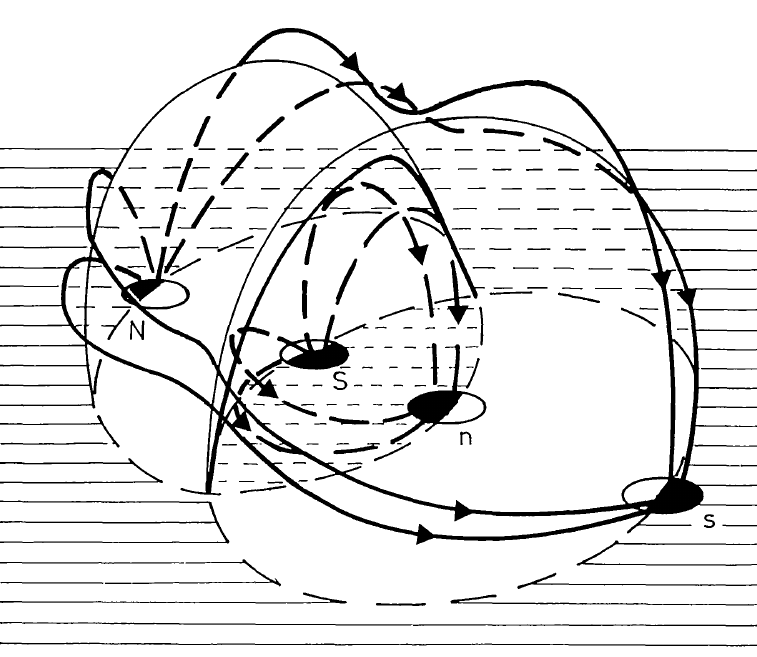}
}
\end{center}
\caption{This cartoon from \cite{1987A&A...185..306H} shows on the lefthand side a simple arrangement of 4 magnetic sources in a line, with the single arrowheads indicating the field and the double arrowheads indicating the current. The solid lines indicate the 2-D separatrix field lines separating 4 domains of different connectivity (including the exterior domain) and intersecting in an X-point. The righthand side shows this in 3-D; the separatrix lines are now domed separatrix surfaces, and the X-point is extended into a separator field line.  This figure begins to suggest the structural complexity that can be arrived at from multiple magnetic sources in 3-D}\label{fig:topology}
\end{figure}

In a 2.5D model like CSHKP, the interface between closed and open field is an example of a separatrix surface. In more realistic topologies, separatrix surfaces curve around and intersect with one another at `separator' field lines. The identification of such flux domains and intersections as locations of particular significance in a flare was made by \cite{1969ARA&A...7..149S}, and developed by many authors. An extensive review can be found in \cite{2005LRSP....2....7L}. Particularly relevant for flares is the work of \cite{1987A&A...185..306H} who discussed how coronal currents could be generated and flow along coronal separators and \cite{1992SoPh..139..105D} who developed an early observationally motivated 3D model of coronal separators and separatrix surfaces (see Figure~\ref{fig:topology}). The calculated positions of the intersections of separatrix surfaces \citep[or, more realistically in the case of continuous photospheric field distributions, quasi-separatrix layers, e.g. ][]{1995JGR...10023443P} with the photosphere agree well with the observed positions of flare ribbons \citep[e.g.][]{1991A&A...250..541M}. Quasi-separatrix layers are locations where coronal `slip-running' reconnection may occur, leading to brightenings running along ribbons \citep{2009ApJ...700..559M}. 
Separators may also be critical structures, though this is by no means so clear. Theoretically, the minimum energy state of the corona consistent with a given photospheric boundary (extrapolated from discrete magnetic charges) is one in which the currents are concentrated along separator field lines while the rest of the field is current-free - the `minimum current corona' model of \cite{1996SoPh..169...91L}. This could be seen as the state to which a corona evolves under slow driving. Observationally \cite{2003ApJ...595..483M} and \cite{2009ApJ...693.1628D} identify some HXR footpoints with the photospheric ends of separators, which move when magnetic flux is transferred between magnetic domains as reconnection proceeds. There are relatively few studies in the literature dealing explicitly and carefully with the magnetic  and topological evolution of a flaring active region and the primary (HXR) emission sites within it,  so more case studies and equally importantly a proper synthesis of these case studies to identify common traits are necessary to understand the significance of the locations of these strongest energy release sites. This will have to involve detailed study of the peculiarities of each event before a coherent picture can be established - a very time-consuming process.

We note that, as expected locations of strong currents, separators, separatrix surfaces and other singular (or quasi-singular) structures have frequently been proposed as sites of particle acceleration. We return to this in Section~\ref{sect:accel}.

\subsection{Magnetic structure, complexity, and flaring}\label{sect:structure}
Substantial effort has been expended on seeking relationships between various properties of sunspot groups and photospheric magnetic fields, and the flare productivity of a region. 
The level of complexity in the active region photospheric field is generally reflected in the complexity of the spatial arrangement of sunspot umbrae and penumbrae
\citep{1990SoPh..125..251M}.  We do know for certain that  large, complicated sunspot groups which are rapidly evolving tend to be flare-productive, but it is likely to be difficult to pick apart the properties of an active region that leads to a propensity to flare - for example rapid emergence (tending to lead to strong magnetic gradients and high shear), a high total flux, or complexity. Each of these is linked to a possible scenario for flare occurrence. Rapid emergence would lead to the development of strong current sheets at the interface with pre-existing coronal field which has been proposed as a trigger \citep{1974SoPh...34..323H}. Strong shear or twisting flows may also indicate the rapid emergence of a  flux rope \citep{2004ApJ...610..588M}, or of a  structure carrying a high current \citep{2003ApJ...586..630M}. A complex field implies a complex topology, with multiple `stress points' - the nulls, separators and QSLs discussed previously at which reconnection is likely to happen

Both global properties (e.g. total magnetic flux and current) and local properties (e.g. variations in the magnetic shear around the neutral line) are important for flaring, but it is surprising how little can be said about flare productivity from photospheric observations: `the state of the photospheric magnetic field at any given time has limited bearing on whether [a] region will be flare productive' \citep{2007ApJ...656.1173L}.  Flaring behaviour also depends on more subtle properties of the magnetic structure. Perhaps the `disconnect' between photospheric and coronal magnetic fields implied by the change from non-force free to force free across the chromosphere means that photospheric fields will never hold the key to flare prediction. Perhaps it is necessary to look in detail at the distribution and driving of topological structures of the magnetic field. Or perhaps it is necessary to understand microphysical properties - for example the development of resistivity in a coronal current sheet - which might never be accessible to our observational or theoretical tools, leaving us in  the disheartening situation that the time and exact location of flare onset are determined by plasma properties of which we have only a weak observational grasp

\subsection{The role of magnetic structures in flare and eruption onset}\label{sect:erupt}
The basic flare structure is a magnetic field carrying strong electrical currents, line-tied at a high inertia photosphere and embedded in a magnetised corona which supports currents with sub-photospheric origins (see Kuijpers et al., this Volume). The manner in which a flare or eruption starts and develops depends both on the intrinsic stability of the strong-current structure and on how it interacts with the surrounding magnetic field, and given the enormous variety of configurations that can arise it may well be that there is not a unique mechanism.

Flares can be eruptive or `confined' (not exhibiting a CME), and one can imagine all sorts of reasons for the distinction. For example, internal reconnections in a loop being twisted at its base and supporting internal currents could lead to flaring energy release inside the loop without any associated ejection, via a (non-erupting) kink instability \citep{2009ApJ...697..999L}. Similar ideas are proposed for micro-flaring \citep[e.g. ][]{2010A&A...521A..70B}. Eruption on the other hand requires rapid inflation of a closed field, e.g. by photospheric current injection (thought unlikely on the timescale of a flare) or by removing or weakening overlying magnetic structure, or by allowing material on closed field access to open field via reconnection. The flare and eruption is thought likely to start with slow evolution towards an MHD instability, followed by the instability and reconnection permitting the magnetic field to reconfigure as the MHD instability dictates. In support of this, filaments are often observed to start to rise slowly, indicating the loss of equilibrium of a previously stable magnetic system, before the main flare thermal or non-thermal radiation is detectable \citep{1988ApJ...328..824K,2005ApJ...630.1148S} 

\begin{figure}
\centering
\includegraphics[height=8cm]{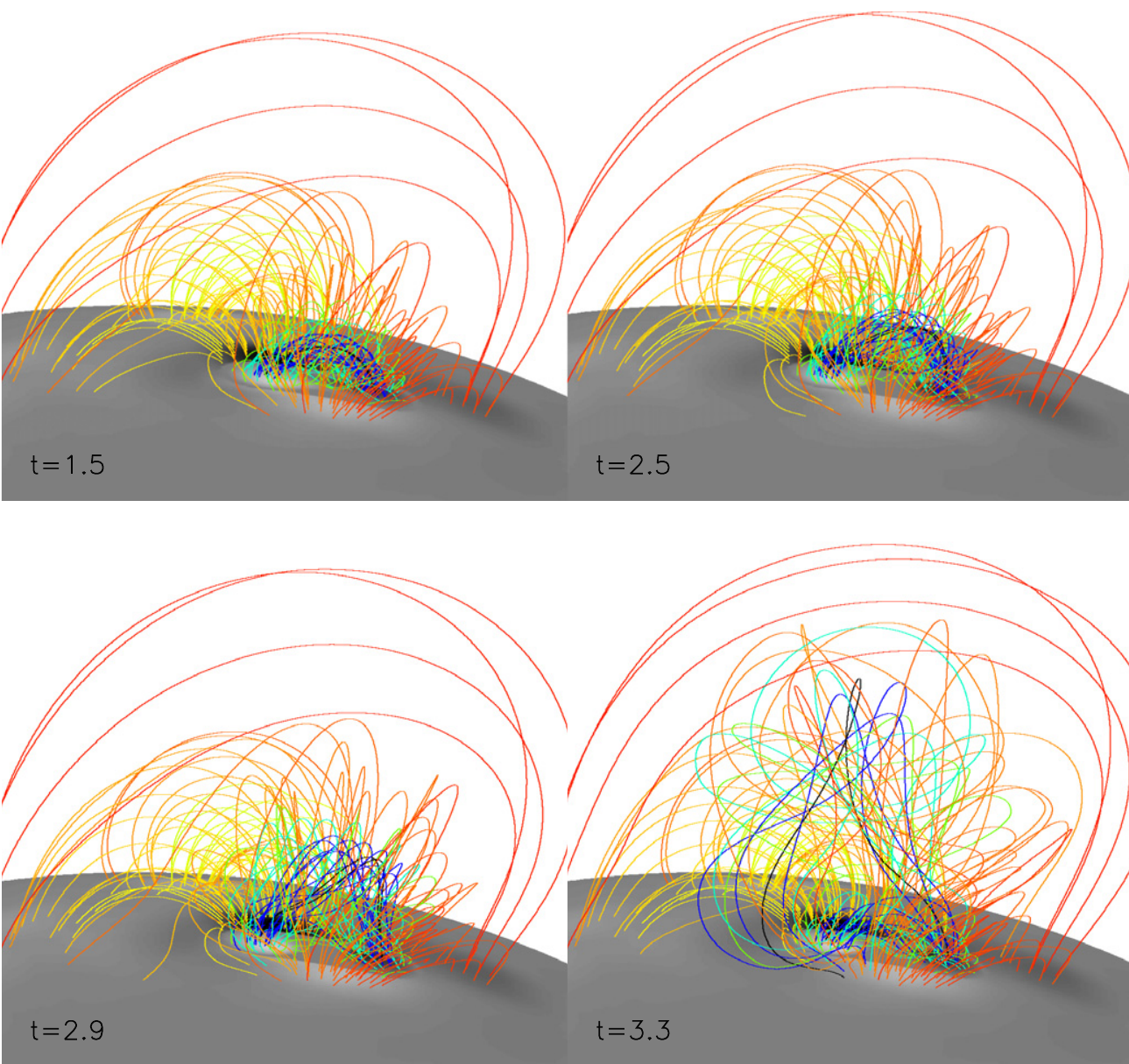}
\caption{Snapshots from an MHD simulation by \cite{2011ApJ...740...68F} of a solar eruption that occurred on 13 December 2006. The fieldlines drawn show a compact flux rope (blue/green lines) embedded in a larger-scale potential field. Just prior to the snapshots shown the flux rope has emerged rapidly through the lower boundary (photosphere) and this is followed by a period of slow emergence, during which the eruption occurs because the rope fails to reach a static equilibrium with its surroundings.} \label{fig:fan_erupt}
\end{figure}

Theory tends to consider two broad classes of coronal structure that can become unstable: sheared arcades and flux ropes.   In the former class, a set of magnetic loops rooted on either side of a polarity inversion line are driven by photospheric shear flows, inflating the field until it erupts \citep{1994ApJ...430..898M}. During the eruption, reconnection between and underneath the loops of the arcade result in the formation of a twisted flux rope which is expelled into space, and the flare occurs as the coronal field re-organises behind this. In the latter class, the initial configuration is modeled as a flux-rope that has either emerged bodily from underneath the photosphere into an overlying arcade field that stabilises it, providing a magnetic tension force to counteract the hoop force in the flux rope \citep{1999A&A...351..707T} or has been formed in situ as reconnection happens during the emergence of a sheared arcade \citep{2003ApJ...585.1073A}. Eruption happens (e.g. Figure~\ref{fig:fan_erupt}) when the flux rope is perturbed by some critical amount leading to one of the classical MHD instabilities of a twisted flux tube \citep{2005ApJ...630L..97T,2006PhRvL..96y5002K,2007ApJ...668.1232F}. 

Interaction of the stressed magnetic structure carrying the free energy for the flare with its environment is central  in understanding what drives what.  For example, in a simple geometry with a rope emerging into an overlying arcade, the flux rope becoming kink-unstable can force the field aside and burst through, with reconnection between flux rope and overlying field resulting from the MHD instability \citep{2007ApJ...668.1232F}. On the other hand in the `magnetic breakout' model \citep{1999ApJ...510..485A} it is reconnection between the energy-storing sheared structure and the surroundings that destabilises the system and leads to the eruption. 
 \cite{2013AdSpR..51.1967S} argue that all of the proposed mechanisms can play a role, but favour the  torus instability \citep{2006PhRvL..96y5002K} that occurs when the outwards Lorentz force of a current-carrying ring or partial ring of magnetic field (the `hoop' force) is unbalanced by forces exerted by an external or `strapping' field. The reasons given by \cite{2013AdSpR..51.1967S} for favouring this instability are that simulations of other types of instability have failed to produce eruptions, and also that observationally the occurrence or otherwise of an eruption seems to depend on the decay of the magnetic field with height: the torus instability is known to require an environment where the overlying field with height more rapidly than some critical value \citep{2007AN....328..743T}.

Adding another complicating factor, the global coronal magnetic field on its largest scales can have an influence on the onset of flares. With high time-resolution and coverage of the complete solar disk from the Atmospheric Imaging Assembly on the Solar Dynamics Observatory it has been established that an event in one active region can be directly (causally) connected to an event in another region ~\citep{2013ApJ...773...93S}. Though not common, it does imply that the structure and stability of the global solar coronal magnetic field, and not just that of the region where the flare occurs, may play a role in enabling or preventing a flare and CME.

\subsection{Flares at different scales and relationship to coronal heating}
Flare parameters (e.g. total thermal and non-thermal energy, peak power) follow power-law distributions, suggesting at least some similarity of structure across physical size scales ~\citep{2011SSRv..159..263H}.  It is clear that  small flares (i.e. microflares, of the smallest GOES classification) do share many characteristics with larger events, such as spatial morphology and production of non-thermal particles \citep{2001ApJ...557L.125L,2008A&A...481L..45H}.  Individual coronal `nanoflares' have not been observed though a response in the lower atmosphere may have been \citep{2013ApJ...770L...1T} and so it is yet to be established whether the low-energy end of the observed flare distribution continues smoothly into these proposed coronal heating events. There is also as yet no evidence that the process that heats the non-flaring corona produces accelerated electrons with the high non-thermal energy content that characterises larger flares \citep{2010ApJ...724..487H}, though this may merely be a detector sensitivity issue. 

As mentioned in Section~\ref{sect:erupt} the kink instability in a strand subjected to twisting motions about its own axis is a model for confined flares, and a model of coronal strands wrapped about one another is also proposed as a mechanism for heating the corona \citep{1988ApJ...330..474P}.  The idea of storing energy for a flare in a realistic corona by Parker-like random shuffling of its photospheric footpoints has also been investigated by  \cite{2013A&A...550A..30B} and by \cite{2011A&A...529A.101D} who incorporated a sandpile-like release. Statistical distributions of flare populations can be obtained in such models, as can the bursty behaviour characterising individual events. In both geometries it is the formation of current sheets and either Joule heating or component reconnection that leads to the coronal heating and energy dissipation, but whereas coronal heating requires a quasi-continuous transfer of energy from field to particles, flares require instead that energy is stored for some time, and released intermittently, in large events. It is not clear why the same twisting or shuffling process should have such different outcomes, but the character of the braiding may be an important factor in determining whether a continuous and space-filling heating arises, rather than a more flare-like intermittent behaviour \citep{2011A&A...536A..67W}.

\subsection{Magnetic structures and flare particle acceleration}\label{sect:accel}
The main energy release phase of a solar flare is characterised by intense bursts of radiation generated by accelerated electrons and ions, i.e. bremsstrahlung hard X-rays, and nuclear line and continuum $\gamma$-radiation. The bremsstrahlung radiation is observed to originate primarily in the solar chromosphere, but coronal HXR sources occur frequently as well \citep{2008A&ARv..16..155K}. Imaging of nuclear  $\gamma$-radiation is much more difficult but a lower atmosphere location has been identified in a small number of strong flares and, curiously, it is not always consistent with the location of the non-thermal HXRs \citep{2006ApJ...644L..93H}.
 There has been no direct imaging of nuclear $\gamma$-ray sources in the corona, but $\gamma$-ray observations from flares with chromospheric footpoints over the limb clearly show evidence for large populations of accelerated coronal ions \citep{2000SSRv...93..581R}.  Detections {\emph{in situ}} in space, and via radio emission, give another view of the particle population and its access to the heliosphere.

There are different roles for magnetic structure in particle acceleration. Charged particles are accelerated by electric fields which can arise from time-varying magnetic fields in MHD waves or turbulence, or in magnetic reconnection regions. Models involving acceleration of particles in the electric field set up in a reconnecting singular structure all face a problem, which is exacerbated as the dimension number of the structure reduces from current sheets (or separatrix layers) to separator lines, to nulls. If this accelerator is physically separate from the chromospheric location where the radiation appears, a very large number of electrons per second is required to explain HXR observations - on the order of $10^{36}$ electrons per second \citep{2003ApJ...595L..97H} for a few minutes. In a (pre-flare) corona of typical density $10^{9}$ electrons~$\rm{cm}^{-3}$ a large volume of coronal plasma per second needs to be `processed' through the reconnecting structure. A single large current sheet of $\sim$30,000~km on a side, with an external Alfv\'en speed of 1000${\rm km~s^{-1}}$ could provide this number flux, but any other single reconnecting structure one can imagine, e.g. an X-line, cannot \citep{2006SoPh..236...59H}. 

\begin{figure}
\centering
\includegraphics[height=6cm]{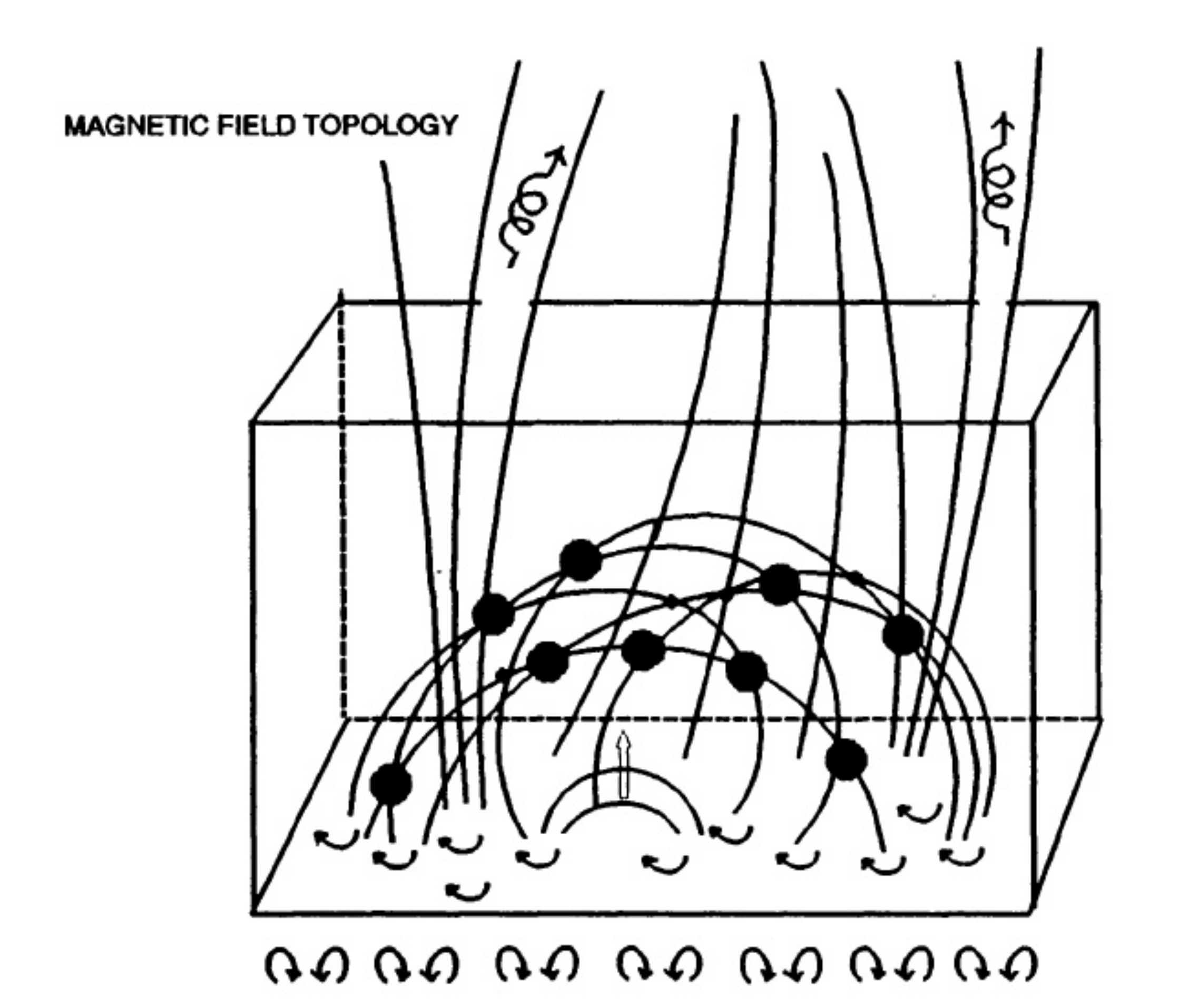}
\caption{A flare cartoon from \cite{1994SSRv...68...39V} showing multiple sites of particle acceleration within a complex coronal magnetic structure. Though the sites are shown here distributed randomly through the corona, they must exist within the overall large-scale magnetic organisation revealed by flare observations and field recontructions.} \label{fig:vlahos_cartoon}
\end{figure}

Acceleration throughout a large coronal volume, in turbulence or by multiple interactions with many smaller current sheets as shown in Figure~\ref{fig:vlahos_cartoon}, is often proposed \citep[e.g.][]{2005ApJ...620L..59T,2012SSRv..173..223C}, though such `volumetric' acceleration still does not fully address the number requirements of a coronal acceleration model. The required electron rate and typical pre-flare coronal density implies the equivalent of $\sim 10^{27} \rm{cm}^{-3}$ of corona being emptied of all electrons each second.  
The separation of impulsive phase HXR footpoints, typically 20-60 arcseconds or 15,000-45,000 km on the surface of the Sun \citep{2008SoPh..250...53S} suggests a coronal volume involved of a few $\times 10^{27}-10^{28}{\rm cm}^{-3}$, which will therefore need to be replenished during the flare. It may be possible to do this by return flows in the electric-field-free parts of the plasma. 
If a means can be found to increase the bremsstrahlung yield per electron (e.g. acceleration/re-acceleration in the radiating source) the demands on electron number or supply can be reduced.
 
Disordered small-scale field structures must exist within a large-scale organising structure, defined by large-scale field stresses before the flare, and ordered post-flare loops afterwards. Individual flare HXR lightcurves show bursty behavior that can be characterised as fractal in time \citep{2007ApJ...662..691M}, but inspection of flare images show that in space the HXR sources undergo rather ordered motions, associated with the evolution of the large-scale magnetic field \citep{2005ApJ...625L.143G}. Whatever is happening in the corona, it is not completely random in an individual flare event.  The observed power-law distributions in the properties of large numbers of flares have led to notions of self-organised criticality (SOC) being applied to individual flares \citep[e.g][]{1995A&A...299..897V}, but the observed distributions show only that the ensemble of active regions over some large portion of a solar cycle exhibits SOC-like scaling \citep{1991ApJ...380L..89L}. It remains to be seen whether SOC ideas can be applied successfully also to an individual flare in a single active region.

\subsection{Discussion}
Understanding flares and eruptions requires a deep knowledge of coronal magnetic structures. The free energy of a flare is stored in non-potential magnetic fields in the low-$\beta$ corona, which emerge from below the photosphere, and the  current-carrying magnetic flux rope is therefore a basic ingredient of the active corona. The current paths through the corona are determined by its topological structure and may be complex. Topology and topological changes also determine the post-flare states that can be accessed (and thus how much free energy can be released), and whether or not a magnetic restructuring will lead to an ejection. Within the larger scale structure, smaller structures must exist or form to allow energy to be transferred to scales at which both thermal plasma and non-thermal particles can pick it up. In more than 40 years of study the theory and observation of many aspects of flares and ejections have become highly refined but the answers to basic questions, such as identifying the conditions that precede a flare, or identifying how and where flare non-thermal particles are accelerated, still elude us.

\section{The open solar corona}

\begin{figure}
\centering
\includegraphics[height=8cm]{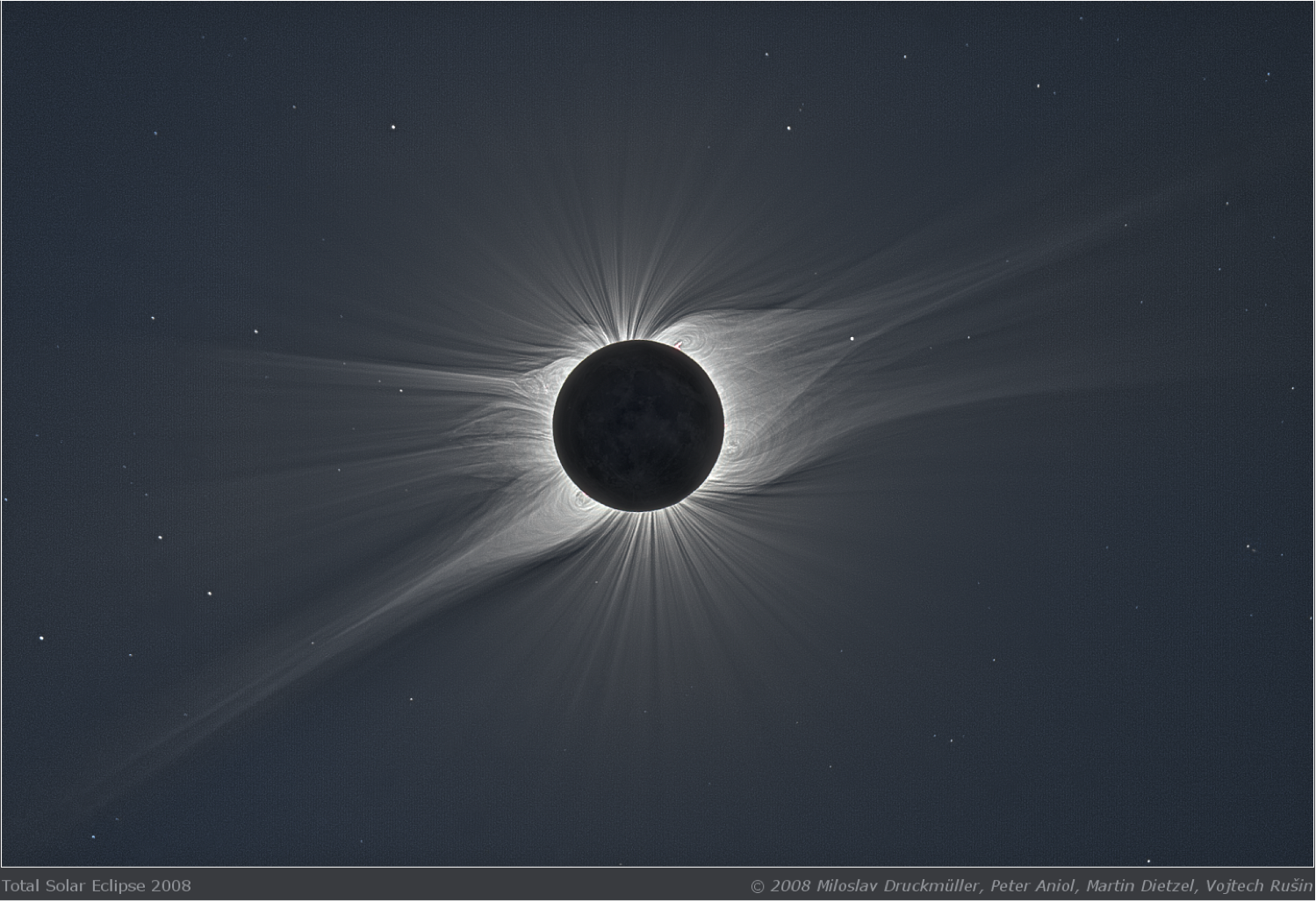}
\caption{White light image of the solar eclipse of Aug 1, 2008 \citep{2009ApJ...702.1297P}} \label{fig:eclipse}
\end{figure}

The study of the Sun's open magnetic field corona began with the \citet{1958ApJ...128..664P} theory of the steady, spherically symmetric solar wind. Parker argued that if the solar atmosphere were static with a temperature of order 1MK out to several solar radii, then the resulting gas pressure at infinity would be finite. Consequently, the corona cannot be static; instead, it must expand outward to interstellar space as a steady, supersonic outflow, as was confirmed a few years later by the Mariner II observations \citep{1962Sci...138.1095N}. However, it is known that the true corona and wind exhibit diverse structure, as shown in the eclipse image shown in Figure~\ref{fig:eclipse}. Note that this image was taken near solar minimum; at solar maximum the structure is even more complex. This spatial complexity is primarily due to the distribution of magnetic flux at the photosphere, which has enormous structure throughout the solar cycle.  Given that the flux distribution at the photosphere is constantly evolving via flux emergence, cancellation, and the multi-scale photospheric motions, the large-scale corona of Figure~\ref{fig:eclipse} must be fully dynamic and unlikely to be in a true steady state. If the photospheric flux evolution is slow, however, then a quasi-steady approximation may be valid. This is the fundamental assumption for most present-day modeling of the large-scale solar/heliospheric magnetic field.

\subsection{Structure and Dynamics of the Open-field Corona}

As with the magnetically closed solar atmosphere, the structure and associated dynamics of the magnetically open corona are due to the effects of the Sun's magnetic field: the freezing of plasma and field implies that the structure seen in Figure~\ref{fig:eclipse} traces out the magnetic field. Figure~\ref{fig:eclipse} also shows two primary types of structure: the finite-length arcs or loops discussed in Section 3 above and the semi-infinite rays associated with open magnetic field lines that extend from the solar surface out beyond the edge of the image, along which the solar wind must flow. While these open field lines extend outward to the heliopause at $\sim 140$ AU, the large-scale properties of the heliosphere, as well as the distribution of open and closed flux, are governed by the magnetic structure and dynamics at the photosphere. For example, the most widely used model for the global coronal magnetic field is the Potential Field Source Surface (PFSS) model, which assumes that in the low corona, below some radius $R_s \sim 2 R_{\odot}$, the field is potential, and purely radial at $R_s$ \citep{1969SoPh....9..131A,1969SoPh....6..442S,1991AdSpR..11Q..15H}. For production models such as the Wang-Sheeley-Arge \citep{2000JGR...10510465A} the source surface radius is held fixed, so that the only real input to the PFSS model is the observed normal flux at the photosphere. Given the extreme simplicity of the PFSS model and the errors inherent in the input data, the model does surprisingly well at reproducing the observed large-scale distribution of open and closed field at the Sun \citep{2006ApJ...653.1510R}, at least, during solar minimum when the photospheric flux is not changing too rapidly. The reason is that $\beta << 1$ in the low corona, and the field is believed to be close to potential except at filament channels, which lie in the innermost regions of the closed field, such as seen in the helmets at the NW and SE (upper right and lower left) limbs in Figure~\ref{fig:eclipse}. Fully 3D MHD models also show that the open flux structure is determined predominantly by the photospheric flux distribution \citep{2006ApJ...653.1510R}. 

In fact there are two types of open field regions in the corona. The most obvious are the large-scale, long-lived open regions corresponding to coronal holes as seen in EUV (Figure~\ref{fig:sdo_overview}) which evolve quasi-statically. The second are the boundaries of these coronal holes, which are likely to be fully dynamic and have a scale of order a supergranule at the photosphere. Note that if we assume a purely radial extrapolation, this scale corresponds to an angular width of order 5 degrees or so in the heliosphere, which is roughly the width of the streamer stalks observable in Figure~\ref{fig:eclipse}.

Accompanying the spatial spectrum of photospheric flux is a spectrum of temporal scales ranging from days for the emergence and disappearance of active region flux to minutes for the elemental flux of the so-called magnetic carpet \citep{1997ApJ...487..424S}. In addition, there exists a complex of photospheric motions such as the granular and supergranular flows that continuously drive the coronal fieldlines at their photospheric footpoints.  All this activity at the photosphere and interior is imprinted onto the corona and heliosphere. Small-scale photospheric motions are expected to result in upward propagating Alfv\'en waves on field lines near the Sun's poles, such as seen in Figure~\ref{fig:eclipse}. These waves have long been proposed as the source of the energy and momentum that powers the wind \citep{2012SSRv..172..145C}, and are likely to be a major source of solar wind turbulence \citep{2010ApJ...708L.116V}. Indeed some authors have argued that the supergranular structure can be seen directly in the heliosphere as well-defined fluxtubes \citep{2008JGRA..113.8110B}; but there is debate over this result \citep{2009ApJ...691L.111G}.

A key question that is only now starting to be explored in detail is the effect of the photospheric motions on field lines near the interface between open and closed flux, such as the boundaries of the various streamers in Figure~\ref{fig:eclipse}. Topologically, the open-closed interface is a separatrix surface of zero width \citep{1990ApJ...350..672L}, but this holds only for a true steady-state. Dynamically, the width of the interface will be set by the time-scale for establishing a steady wind, generally a day or so, which for typical photospheric speeds $\sim$ 1 km/s corresponds to a spatial scale of $\sim$ 30,000 km, the scale of a supergranule. For time/spatial scales longer than this the open-closed interface can be considered to evolve quasi-steadily, well approximated by, for example, a sequence of PFSS solutions. On smaller scales, however, the interface must be fully dynamic and the opening and closing of field lines calculated explicitly. 

\subsection{Fast and slow solar wind}
\begin{figure}
\begin{center}
\hbox{
\includegraphics[height=6cm, width=0.45\textwidth]{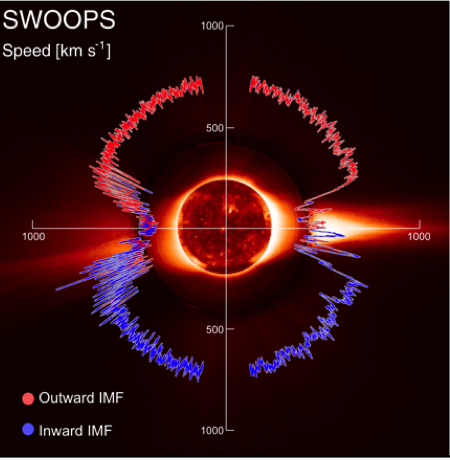}
\includegraphics[height=4cm]{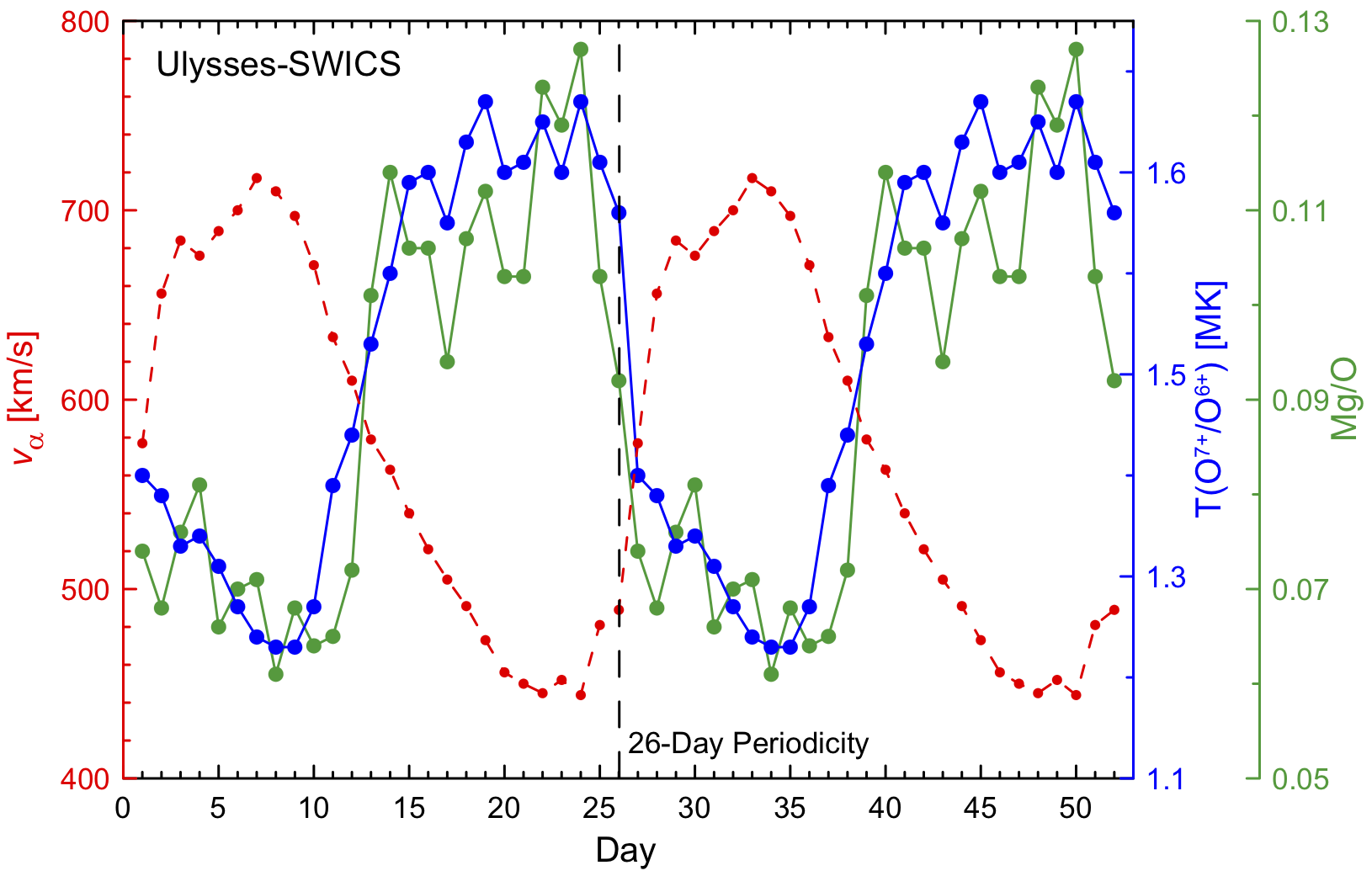}
}
\end{center}
\caption{Left Panel: Solar wind speed versus latitude from \citet{2008GeoRL..3518103M}. The inset consists of a SOHO/EIT image imbedded in a SOHO/LASCO C2 image. Right Panel: Solar Variation of solar wind speed (red), charge state (blue) and composition (green) for a time period during the 1995, from \citet{1995SSRv...72...49G}.}\label{fig:ulysses}
\end{figure}

In parallel with the two types of open field regions in the corona, it has long been known that there are two distinct types of solar wind: the fast and slow. This can be seen in the left panel of Figure~\ref{fig:ulysses} \citep{2008GeoRL..3518103M}, which shows the speed of the wind as a function of heliospheric latitude between 1992 and 1998. These measurements were obtained during solar minimum, when the large-scale photospheric magnetic field is predominantly dipolar and exhibits the least global structure and evolution. Note, however, that due to the constant emergence and cancellation of the small-scale dipoles of the magnetic carpet, the small-scale field of the chromosphere and corona is always far from a steady state. It is evident from the Figure that the wind at high latitudes is fast, speeds $>$ 500 km/s, and fairly steady, while that at low latitudes is slow $<$ 500 km/s, with large variability. Note that the slow wind surrounds the region at the ecliptic where streamer stalks and the heliospheric current sheet (HCS) are located. 

This extreme dichotomy of the two winds suggests that they have different sources at the Sun and is best examined by measuring the compositions of the two winds, because the plasma composition directly connects the wind material to its source at the Sun \citep{2002GeoRL..29.1352Z}. Indeed speed is not a valid discriminator of the two types of solar wind. The wind from small equatorial coronal holes is often observed to be slow, V $\sim$ 500 km/s, but yet it has the spatial, temporal, and compositional signatures of the fast wind \citep{2009GeoRL..3614104Z}. 

To demonstrate this, the right panel of Figure~\ref{fig:ulysses} shows a superposed epoch analysis of solar wind composition for ten solar rotations during 1992 - 1993 when Ulysses was periodically sampling fast and slow wind during a solar rotation \citep{1995SSRv...72...49G}. The red curve plots the alpha particle speed, which is similar to that of protons, and shows the characteristic smooth gradient where the fast stream outruns the slow and a sharp gradient where the fast runs into the slow. The blue curve shows the freeze-in temperature of the plasma back near the Sun as derived from charge state ratios of Oxygen ions. There is a clear correlation with wind speed; the slow wind originates from markedly hotter plasma than the fast. The green curve plots the ratio of Magnesium, an element with low first ionization potential (FIP) to Oxygen, which has high FIP. Note that the ratio is consistently a factor of 3 or so higher in the slow wind than in the fast. The FIP ratio of the slow wind is very similar to that measured for the closed corona, while that of the fast wind is similar to that measured for the photosphere.

From these and many other observations of the wind and corona, the differences between the two winds can be summarised as follows:\\
{\bf (1) Spatial Properties:} The fast wind is predominantly found at high latitudes near solar minimum and, consequently, is believed to originate from long-lived coronal holes, as can be seen from Figure~\ref{fig:ulysses}. The slow wind is found at low latitudes and surrounds the HCS (heliospheric current sheet) \citep{2002JGRA..107.1410B}.  The HCS is always observed to be embedded in slow wind, which is sometimes observed to extend as much as $30^{\circ}$ or more from the HCS.\\
{\bf (2) Temporal Properties:} The fast wind has predominately steady speed and composition \citep{1995SSRv...72...49G,1995AdSpR..15Q...3V,2007ARA&A..45..297Z}, similar to the quasi-steady wind of \citep{1958ApJ...128..664P} but with a significant amount of additional physics such as momentum deposition, turbulence, and kinetic effects \citep{2007ApJ...662..669V}. The slow wind, on the other hand, is strongly variable in both speed and composition \citep{2007ARA&A..45..297Z}.\\
{\bf (3) Composition:} The ionic composition of the fast wind implies a freeze-in temperature at its source of  $ T \sim 1.2$ MK, typical of coronal holes, and an elemental abundance close to that of the photosphere \citep{2001SSRv...97..123V}. The slow wind has a freeze-in $T \sim 1.5$ MK and an abundance close to that of the closed corona \citep{1995SSRv...72...49G}. Many have argued that composition rather than speed is the physical property that defines the wind, because the wind from small, long-lived coronal holes can be slow ($\sim 500$ km/s) yet has ``fast" wind composition \citep{2009GeoRL..3614104Z}.

At present the source of the fast wind at the Sun is generally accepted to be coronal holes. Both imaging and in situ data support this conclusion. Almost immediately after the initial discovery of coronal holes by Skylab, it was inferred that these are the source of the so-called high speed streams \citep{1977RvGSP..15..257Z}. Since then, many in situ observations of high-speed streams have indicated that they are magnetically connected back to coronal holes on the disk. The Ulysses results are especially definitive \citet{2008GeoRL..3518103M}.  There is no doubt from Fig. 13 that when Ulysses is over a polar coronal hole it sees fast wind. Furthermore, direct spectroscopic imaging of coronal holes appears to show outflows along network boundaries \citep{1999Sci...283..810H}, exactly as would be expected for the source of the fast wind. This result has been further confirmed and elaborated by \citep{2005Sci...308..519T} who used the spectroscopic date to determine the height at which the fast wind starts to flow in the network fluxtubes. The main questions regarding the fast wind, therefore, are not as to its source region on the Sun, but as to the actual mechanism for its acceleration.

\subsubsection{Models for the sources of the slow wind}

Unlike the fast wind, the source of the slow wind remains as one of the outstanding questions in solar/heliospheric physics. Three general types of theories have been proposed for the location of slow wind origin at the Sun.  Given its observed location in the heliosphere and its association with the HCS, as evident in Figure~\ref{fig:ulysses}, all three theories involve the closed field regions in some manner. The three theories differ most strongly in the location of the source regions at the Sun and in the role of dynamics.

{\bf The Expansion Factor Model}. Perhaps the most straightforward theory for the slow wind is the so-called expansion factor model in which the slow wind originates from coronal hole regions, just like the fast wind, but only from open flux tubes near the boundary of the coronal hole with the closed flux region \citep{1979SSRv...23..159S,1988ApJ...325..442W,1990ApJ...355..726W}.  The model was originally described in terms of a single parameter, the expansion factor, defined as the ratio of the area of a flux tube at the PFSS source surface to its area at the solar surface. The basic idea underlying the model is that open flux tubes deep inside a coronal hole expand outward approximately radially, whereas flux tubes very near the boundary expand super-radially due to the presence of the closed flux. For example, the open flux that connects to the Y-point at the top of a helmet streamer has, in principle, infinite expansion. This streamer topology is discussed in detail below. It is well known that a large expansion factor can lead to slower velocities in the usual steady-state wind equations \citep{1980JGR....85.4665H}.

A problem with the original expansion factor model is that it predicts that the wind from the vicinity of pseudostreamers, which are also discussed in detail below, should be fast \citep{2007ApJ...658.1340W}, but instead this wind is observed to be slow \citep{2006ApJ...653.1510R,2012SoPh..277..355R}. Consequently, the expansion factor model has been generalized substantially in recent years to consider the effects of the detailed variations of flux tube geometry on solar wind properties. The speed and other properties of the wind are sensitive to the locations of the heat and momentum deposition in an open flux tube, which in turn can vary greatly with flux tube geometry. In fact, \citet{2007ApJS..171..520C} have shown that non-steady solar wind solutions exist even for completely time-independent flux tube geometry and energy/momentum input. Consequently, the observed differences between the two winds may arise solely from the geometrical difference between open-field flux tubes near some open-closed boundary and flux tubes deep in the coronal-hole interior. It is clear from this discussion that, at least, for the expansion factor models, the exact topology of the open-closed boundary and neighboring open flux is essential for understanding the slow wind.

{\bf The Interchange Model}  Another prominent theory for the sources of the slow wind is the interchange model \citep{1998SSRv...86...51F,2005ApJ...626..563F}, in which the small-scale dynamics of the photospheric magnetic field (e.g., the magnetic carpet) play the central role. This model can be thought of as the exact opposite of the expansion factor model in that the slow wind source is the closed field region and dynamics are all-important. The key idea underlying the interchange model is that open flux is conserved and diffuses via reconnection throughout the corona, even into the apparently closed field regions \citep{2003JGRA..108.1157F}.  Unlike the expansion factor model, which assumes a quasi-steady field, the magnetic field is inherently dynamic and the slow wind is postulated to escape from the closed-field region via continuous interchange reconnection between open and closed flux. The interchange model has obvious advantages in accounting for slow wind observations: it naturally produces a continuously variable wind with closed-field plasma composition, located around the HCS but with large extent. The primary challenge for the model is to verify that interchange reconnection induced by photospheric dynamics does, indeed, enable open flux to penetrate deep into closed-field regions. The simulations, to date, have found that the open-closed boundary remains smooth and topologically well defined, even during interchange reconnection \citep{2009ApJ...707.1427E,2011ApJ...731..110L}.  On the other hand, these calculations lacked the topological complexity of observed photospheric flux distributions; consequently, it remains to be seen whether interchange reconnection with sufficiently complex magnetic topology at the open-closed boundary can produce an effective diffusion of the open flux deep into the closed. We note that for this model, as well, the open-closed boundary topology plays a central role.

{\bf The Streamer Top / S-Web Model} Given its associations with the HCS, its variability, and especially its composition, the conjecture that the slow wind is due to the release of closed field plasma onto open field lines, as in the interchange model above, seems promising. Many authors have argued that this release would naturally occur at the open-closed boundary of streamer tops  \citep{1996JGR...10119957S,1999JGR...104..521E,2004ApJ...603..307E,2005ApJ...633..474R}. This streamer-top model can be thought of as intermediate between the expansion factor and interchange in that the location of the slow wind source is at a boundary region between open and closed, as in the expansion factor, but dynamics are essential, as in the interchange.  The release of closed field plasma at streamer tops can account for all three properties of the slow wind, except for the observation that it can extend up 30$^o$ from the HCS. As discussed above, the expected angular width of the interface region due to supergranular flow is only 5$^o$ or so, which is too small to explain the observations. In recent years, however, it has been realized that for the photospheric flux distributions observed on the Sun, the open-closed boundary involves topological structures such as pseudo-streamers, which add much more topological complexity to the open field than expected from only the streamer belt. This topological complexity, referred to as the S-Web model \citep{2011ApJ...731..112A,2012SSRv..172..169A} results in the existence  of open flux located far from the HCS in the heliosphere, but mapping very near the open-closed boundary back in the corona. Interchange reconnection of such flux with closed field would readily release closed field plasma far from the HCS in the heliosphere. 

A key point here is that the topological complexity of the open-closed boundary invoked by the S-Web is present in the purely steady models, such as a PFSS solution. It is not an assumption of the S-Web model. The invoked dynamics to release closed plasma is, indeed, an assumption, but the large-scale magnetic topology is an inherent feature resulting from the observed photospheric flux distributions and, hence, must be included in any model of the corona and heliosphere. It is clear, therefore, that a starting point for any theory of the origins of the slow wind and of the corona-heliosphere connection, in general, is an understanding of the topology open-closed boundary. In the following section we describe the salient features of this topology.

\subsection{The Open-Closed Boundary and its Extension into the Heliosphere}

For a completely general photospheric field, the open-closed boundary can have arbitrary topological complexity, but for the flux distributions actually observed on the Sun, there are only three types of structures whose boundaries are of importance: streamers, plumes/jets, and pseudo-streamers (originally defined by \citet{1972PCS.....5.....H} as plasma sheets), and all are evident in Figure~\ref{fig:eclipse}. The helmets and stalks on the lower left and right side are streamers; the bright rays emanating from the two poles are plumes (some may be jets); and the bright helmet and stalk on the upper left is a pseudo-streamer. Note also that the three open-field structures appear to originate at different heights above the photosphere: the streamer stalks at $R_s$, the pseudo-streamer at $R_s/4$ \citep{2007ApJ...658.1340W}, and the plumes essentially at the photosphere. The magnetic topology of these boundaries, and associated dynamics, especially magnetic reconnection, are the key to understanding the corona-heliosphere connection. 

\subsubsection{Streamers}
\begin{figure}
\begin{center}
\hbox{
\includegraphics[height=5cm, width=0.45\textwidth]{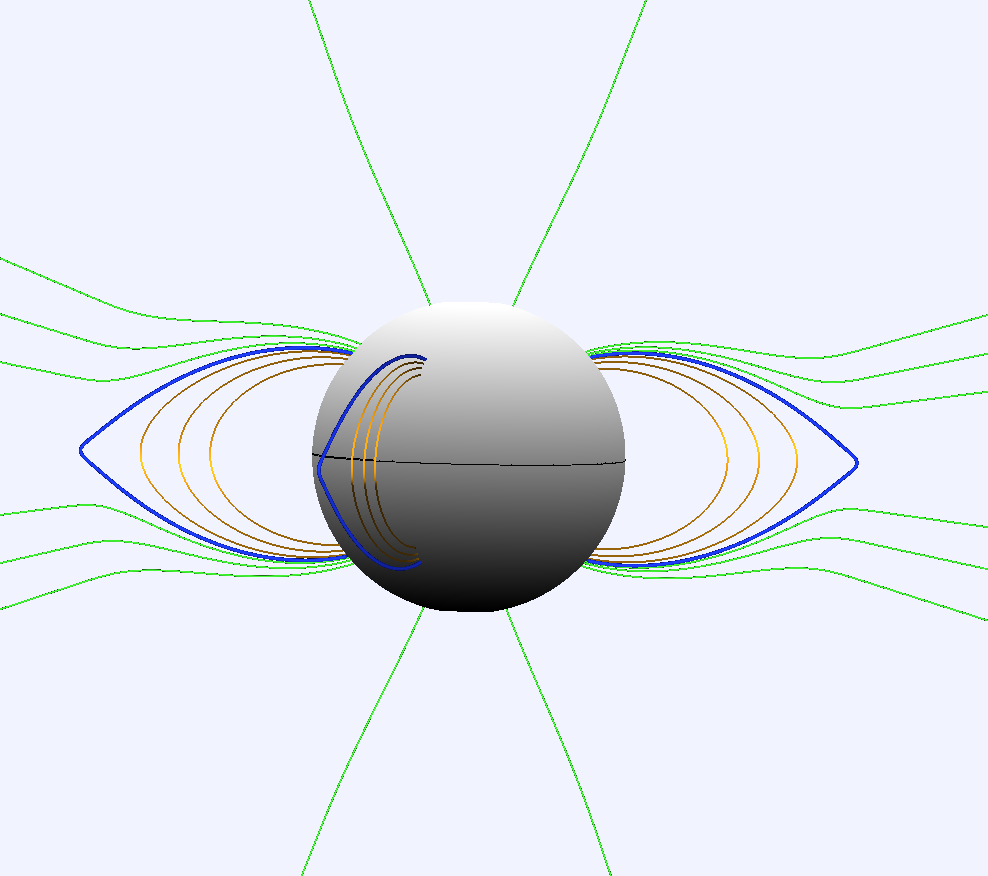}
\includegraphics[height=4cm]{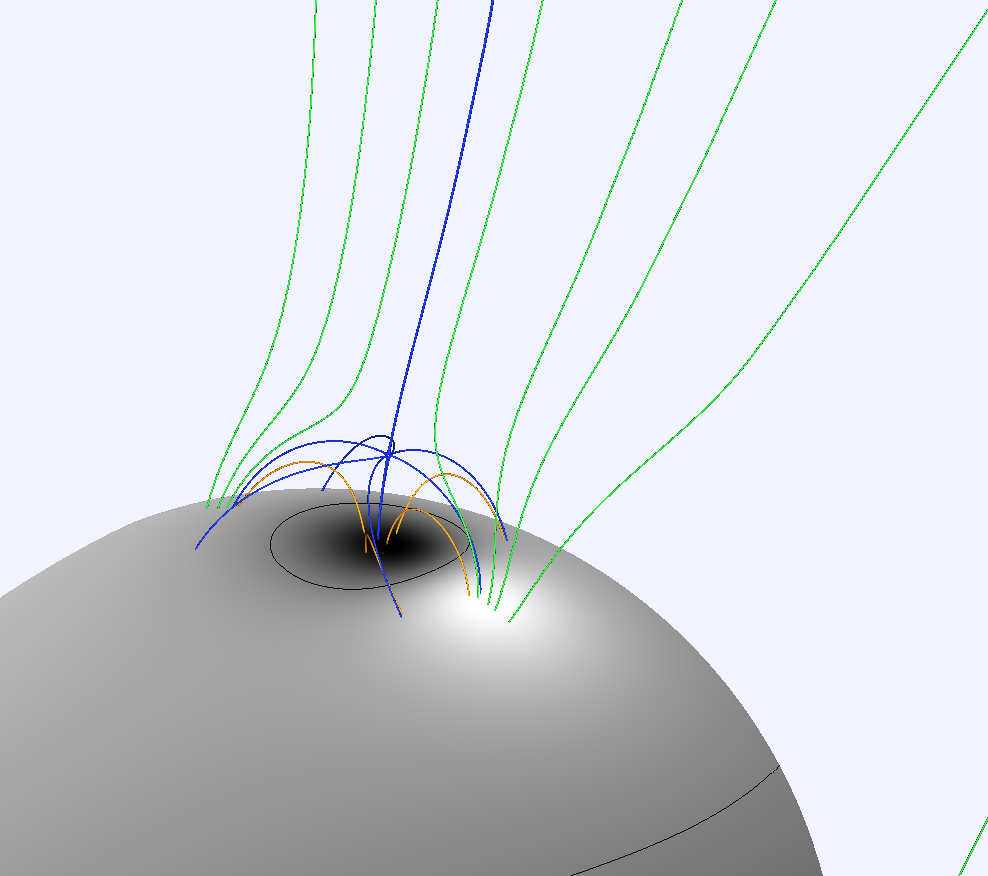}
}
\end{center}
\caption{Left Panel: Source surface model for a helmet streamer topology. Black line on the photosphere is the polarity inversion line at the equator and the field lines in the corona are colored: red for closed, green for open, and blue for separatrix boundary. Right Panel: Source surface model for parasitic polarity region embedded in polar coronal hole of left panel. The blue separatrix lines converge onto a 3D null point and define a fan surface and two spines, one open and one closed. }\label{fig:streamer}
\end{figure}

Assume the simplest possible photospheric flux distribution; that due to a single dipole $\bf d$ located at Sun center. Then the source-surface magnetic field is given by ${\bf B} = - \nabla\phi$, where the potential $\phi$ is,

\begin{equation}
\phi =  {\bf d \cdot r} (R_s^3 - r^3) / (R_s r)^3
\end{equation}

\noindent
and $R_s$ is the radius of the source surface. This field is plotted on the left of Figure~\ref{fig:streamer}. The topology consists of a closed arcade centered about the equator (red field lines) and two polar coronal holes (green open field lines). The boundary between the open and closed regions in the corona is a toroidal-like surface, (the helmet), that wraps around the Sun (blue lines). The intersection of this surface with the photosphere forms two closed curves encircling the Sun roughly latitudinaly that define the boundary between open and closed flux at the photosphere; these are the coronal hole boundaries.  Note that the open-closed boundary is a true separatrix in that the magnetic connectivity is discontinuous across this surface. Consequently, every field line on this surface must end up at a null point. For the simple potential field above, the nulls are of the X-type and form a closed circle that defines the apex of the helmet. These can be seen as the tips of the blue field lines. In the solar case the X-points are deformed by the solar wind flow into Y-points that form a circle, with the connected HCS emanating from the circle. This simple picture agrees well with the helmet streamers observable in Figure~\ref{fig:eclipse}.  

\subsubsection{Plumes/jets}
Let us now consider the simplest possible extension of the magnetic topology above by adding a small dipole $\bf d_1$ below the photosphere. Such a field would correspond to, for example, a single bipolar active region or a single bipole of the magnetic carpet. If the dipole occurs at low latitudes where the coronal field is closed, then it would not change the basic structure of the open-closed topology, only its detailed shape. However, if the dipole occurs in the open field region a new open-closed boundary must occur on the Sun, because a parasitic polarity region (polarity opposite to that of the coronal hole) will appear on the photosphere, and this parasitic flux must be closed. The resulting topology is shown in the right hand panel of Figure~\ref{fig:streamer}. This field is calculated using a source surface model in which a contribution  $\phi_1$ due to the small dipole is added to the potential above: 

\begin{equation}
 \phi_1 = {\bf d_1 \cdot (r - r_d)} / |{\bf r - r_d}|^3   -  R_s  r_d^3 {\bf (d_1 \cdot} (R_s^2 {\bf r} -  r^2 {\bf r_d}) / |r_d^2 {\bf r} - R_s^2 {\bf r_d}|^3 
\end{equation}
where $\bf r_d$ is the location of the new dipole. The new open-closed topology is that of the well-known embedded bipole consisting of, in the photosphere: a closed circular curve defining the boundary between the small closed-field region and the surrounding open, in the corona: a dome-like fan surface with a single 3D null point near the apex of the dome (blue lines in Figure~\ref{fig:streamer}), and in the heliosphere: a single line, the spine, emanating outward from the null \citep{1998ApJ...502L.181A}. There is also a downward spine, but this is simply part of the closed-field system. 

Adding photospheric motions to this topology is expected to produce continuous reconnection at the null and fan surface between the closed flux and the surrounding coronal hole open flux. In this case the release of closed field plasma can occur very low in the corona, $<$ 10,000 km, depending on the size of the dipole. This type of interchange reconnection has been proposed as the mechanism for a broad range of observed phenomena, including coronal jets \citep{2009ApJ...691...61P}, plumes \citep{1998ApJ...501L.217D}, and even the quasi-steady fast wind itself \citep{1992sws..coll....1A}. Since the magnetic carpet is ubiquitous throughout the Sun, we expect coronal holes to be riddled with small closed parasitic polarity regions. Figure~\ref{fig:eclipse} shows the presence of numerous plumes in the polar hole regions, each plume is believed to have a small parasitic polarity region at its base. 

There remain questions, however, in reconciling the model with in situ measurements. Heliospheric measurements of coronal hole wind, the fast wind, do not show evidence for plume structure; nor do they show evidence of closed field plasma that has been released by interchange reconnection. As implied by Figure~\ref{fig:ulysses}, the wind from coronal holes appears to be uniformly fast wind, with no evidence of the variability or composition of the slow. There seems to be a disconnect, therefore, between the heliospheric data and coronal observations and with the dynamic interface model. This disconnect is one of the major puzzles in coronal/heliospheric physics. The Solar Probe Plus and Solar Orbiter Missions will hopefully resolve this puzzle by measuring the wind much closer to its origin at the Sun.

\subsubsection{Plasma Sheets/Pseudostreamers}
The topology of the third bright open-field structure in Figure~\ref{fig:eclipse}, the plasma sheet or pseudo-streamer visible on the North-East limb, can be thought of as an intermediate case between the streamer and plume topology. A high-lying Y-null circle characterizes streamers, whereas a very low-lying X-point characterizes plumes. For a plasma sheet/pseudostreamer the corresponding coronal singularity is a finite line segment. In principle, this could be an actual null-line, but such lines are topologically unstable, so the singular line segment is due instead to a separator connecting two or more null points \citep{1990ApJ...350..672L}. It is well known from studies of the Earth's magnetosphere that separators are equivalent to null lines in terms of sites for reconnection \citep{1988JGR....93.8583G}. 

\begin{figure}
\centering
\includegraphics[height=8cm]{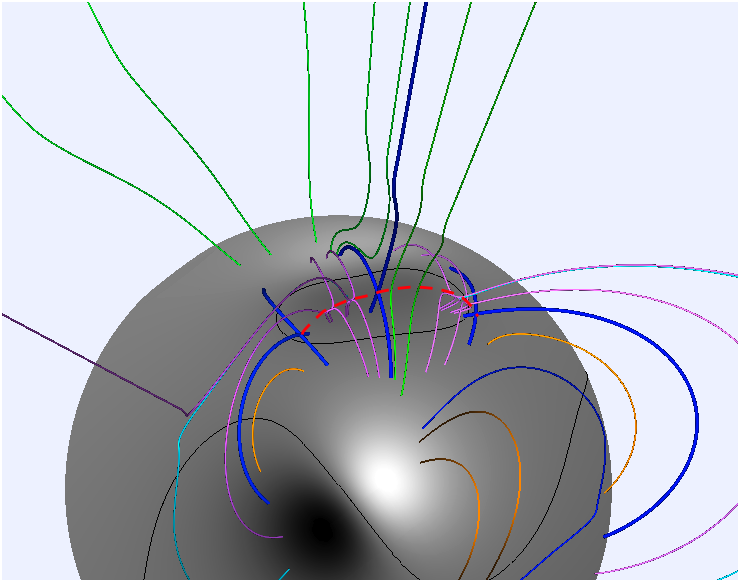}
\caption{Source surface model for a plasma sheet topology. As in Fig. 4, the black elliptical curve at the photosphere is the PIL for the parasitic polarity region, which is inside the coronal hole. The green lines are open, and the intersections of the three pairs of heavy blue lines defines the locations of the three coronal nulls.} \label{fig:plasmasheet}
\end{figure}

Figure~\ref{fig:plasmasheet} shows one source surface model realization of a plasma sheet topology, but we emphasize that this is not unique. Plasma sheets can form from both finite width \citep{2007ApJ...671..936A,2011ApJ...731..112A} and singular width \citep{2011ApJ...731..111T} structures in the corona. Here we will describe only the key features of the singular width case so as to more easily make the comparison with plumes and streamers. The most intuitive picture for the plasma sheet magnetic structure is to start with the embedded bipole of Figure~\ref{fig:streamer}, and then elongate the parasitic polarity region by simply adding more dipoles aligned in a row below the solar surface. This was the procedure used to obtain Figure~\ref{fig:plasmasheet}. If the parasitic polarity region is elongated sufficiently, the coronal null in Figure~\ref{fig:streamer} splits up to form three nulls connected by separator lines. These three points are located at the intersection of the thick blue lines in Figure~\ref{fig:plasmasheet}. Connecting these points are two field lines, the separators, indicated by the red dashed lines. Since these are singular lines it is impossible to find them with a graphics program, so they are simply drawn by hand.

The topology of Figure~\ref{fig:plasmasheet} may appear to resemble that of a plume/jet above, but the nature of the singularity in the heliosphere is very different. In the plume topology above, the singularity is a single line, but for a plasma sheet it is a fan-like surface that emanates from the central null point. The central null has as its separatrix fan a vertical surface that is part open and part closed, while the two side nulls share the dome-like separatrix surface surrounding the parasitic polarity region as their fans. The intersection of the vertical fan of the central null and the dome fan of the side nulls defines the two separator lines where reconnection can occur.

We conclude, therefore, that the open-closed topology of a plasma sheet consists of in the photosphere: a closed elliptical curve defining the extent of the parasitic polarity flux, in the corona: a dome-like surface with multiple 3D null points and separator lines connecting them, and in the heliosphere: a surface of finite angular extent that connects down to the central null and separator lines.  Again, adding photospheric dynamics to this picture implies interchange reconnection all along the separator lines in the corona and the release of closed field plasma into the heliosphere. Note that the release in this case is intermediate to the two cases above, it is not along a single line as in a plume, or along a full planar sheet as in a streamer/HSC, but along a surface that forms only a finite arc in the heliosphere. Furthermore, the height of the release is expected to be intermediate to plumes and streamers in that to have a sufficiently elongated geometry, the parasitic polarity region must be fairly large, but will not be larger that a streamer \citep{2007ApJ...658.1340W}. 

The question remains, however, if pseudostreamers will be a source of slow wind observed in the heliosphere. Plumes apparently are not, but streamers definitely are. This question requires detailed calculations, which have yet to be performed, of the interchange dynamics and closed plasma release for the topology of Figure~\ref{fig:plasmasheet}. Another important question is the amount of slow wind expected from such structures and whether it can account for all the slow wind observed. It may seem that multiple null points with separator lines would form only rarely on the Sun, but in fact, they are fairly common \citep{2012ApJ...759...70T}. As a result of the so-called ``rush to the poles" of trailing polarity flux during the solar cycle, long tongues of opposite polarity often cut into coronal hole regions, leading to structures such as shown in Figure~\ref{fig:plasmasheet}. 

For the observed solar flux distributions the magnetic topology is more complex than described above in that there is inevitably an interaction between the open-closed separatrix surfaces defining the coronal hole and the separatrices of the parasitic polarity region. Although not easily apparent, this interaction is present in the field of Figure~\ref{fig:plasmasheet}. Note that the dark blue lines defining the side nulls are closed, whereas the line defining the central null is open; consequently, the dome fan surface of the parasitic polarity must intersect the open-closed surface of the coronal hole. As a result the heliospheric separatrix surface emanating from the central null and separators takes the shape of an arch whose footpoints lie on the HCS. This is a key result. It implies that regions of slow wind in the heliosphere generally connect to the HCS, but can extend to high latitudes, in agreement with observations. Furthermore, the number of such arches can be so large for an observed photospheric flux distribution that they form a dense web, the S-web, in the heliosphere \citep{2011ApJ...731..112A, 2012SSRv..172..169A,2002JGRA..107.1028C}.

\subsection{Discussion}
We conclude from the discussions above that the nature of the boundary between open and closed field in the corona plays the dominant role in determining the properties of the solar wind, in particular the slow wind. The large-scale geometry of this boundary is determined by the distribution of magnetic flux at the photosphere, but since this boundary is a singular surface, its dynamics are determined by its detailed topology and by magnetic reconnection across this topological singularity. Fortunately, there are only three important cases to consider, streamers, plumes, and plasma sheets, each of which can be seen in eclipse or coronagraph images.  The Sun and Heliosphere, therefore, constitute an amazing example of a multi-scale coupled physical system in that structure at the photosphere on the scale of a solar radius ($\sim 10^{11}$ cm) couples to reconnection dynamics in the corona at kinetic scales ($\sim 100$ cm), which then produce the plasma structures that we observe out at 1 AU ($\sim 10^{13}$ cm)!  

Another major challenge remains the acceleration mechanism for the fast wind. Since there have been many recent reviews of progress on this problem \citep{2012SSRv..172...89H}, we have only only touched upon it here. Two general types of mechanisms have been studied: Alfv\'en waves \citep{2007ApJS..171..520C} and reconnection as in the plume model above \citep{1992sws..coll....1A,2003JGRA..108.1157F}. Most of the work has focused on the wave model, and there are now highly detailed calculations as to how waves generated by photospheric motions can couple nonlinearly to produce turbulent heating and acceleration on 1D open flux tubes \citep{2012SSRv..172..145C,2014ApJ...782...81V}. The challenge is to extend this work to 3D and to include self-consistently the energy cascade down to the dissipation scale in the global models. There has been comparatively much less quantitative work on the reconnection model, because this inherently requires calculation of reconnection dynamics in a fully 3D topology. It appears inevitable, therefore, that further progress on both the slow and fast winds will require a much deeper understanding of multi-scale coupling and reconnection in the Sun's corona.

\section{Conclusions}

This paper has examined the properties of structures in the outer solar atmosphere. For reasons of space, the focus has been on a limited range of topics that have seen progress in recent years. It is clear that many factors have contributed to the major advances seen in the last decade or so. Of these, comprehensive spatial and spectral coverage of the outer solar atmosphere is probably the most significant since it has revealed the complexity of the magnetic and plasma structures, and their inherent dynamics. In turn, this has led to general acceptance that coupling between a hierarchy of scales is central to what is seen and that magnetic reconnection is the prime mechanism driving the observed global phenomena.

At this time, those studying the solar atmosphere have a greater range of data than previously. Putting these different data sources together has yielded outstanding insights, as we have discussed. However, we emphasise the need to undertake extensive parameter surveys: as an example we contrast the results discussed earlier of \cite{2011ApJ...734...90W} who studied a single AR, with those of \cite{2012ApJ...759..141W} who studied almost 20 ARs and reached different conclusions about the time-variability of coronal heating. Such extended studies may not have the excitement of a brand new result, but are essential for understanding the generality and breadth of the result. Unfortunately today due to the very richness of data, there is sometimes a tendency to show just one example and move on.

Numerical simulations also play a major role. For one-dimensional hydrodynamic models as discussed in Section 3, the important length and time scales are now being resolved adequately. For 3D MHD the situation is rather different in that one is dealing with transport coefficients orders of magnitude larger than exist in reality, and this impacts in particular magnetic reconnection studies. For forced (or driven) reconnection, one can argue that numerical reconnection handles the process adequately, at least in a time averaged sense, so that confidence can be maintained in models of large scale eruptions. For smaller scales, and weakly stressed systems, such confidence may be misplaced.

With the launch of the IRIS satellite, the present rush of solar missions takes a pause, until Solar Orbiter in the latter part of this decade and then Solar Probe and Solar-C in the early 2020. The mission instrumentation is largely decided and in some cases under construction: Solar Orbiter will study the connection between the solar surface and solar wind at 0.3 AU, Solar Probe will fly closer to the Sun (8.5 $R_s$), and Solar-C will fly next-generation coronal spectrometers and imagers, as well as measuring the magnetic field higher in the chromosphere. Beyond these, what else would be important to do? The solar flare problem is unsolved. We do not know how to accelerate the required number of particles. So a ```son of RHESSI"' with improved instrumentation, perhaps along the lines of the FOXSI rocket flight, and complementary imaging spectroscopy, is desirable. 

\begin{acknowledgements}
We thank Andre Balogh for organising this workshop and ISSI staff for their hospitality. The work of L. Fletcher has been supported by STFC grant ST/I001808/1 and that of S. Antiochos has been supported by the NASA  TR\&T and SR\&T Programs.
\end{acknowledgements}

\bibliographystyle{ssrv}


\end{document}